\documentclass[nofootinbib,amssymb,amsmath,preprintnumbers,floatfix,aps, 12pt]{revtex4}
\usepackage{graphicx,bm,multirow}
\usepackage{hyperref}
\usepackage{xcolor}   

\usepackage{lipsum}

\makeatletter
\renewcommand\frontmatter@abstractwidth{\dimexpr\textwidth-0.2cm\relax}
\makeatother

\hypersetup{
    colorlinks=true,       % false: boxed links; true: colored links
    linkcolor=red,          % color of internal links
    citecolor=blue,        % color of links to bibliography
    filecolor=red,            % color of file links
    urlcolor=black,              % color of external links
}

\newcommand{\be}{\begin{equation}}
\newcommand{\ee}{\end{equation}}
\newcommand{\bear}{\begin{eqnarray}}
\newcommand{\eear}{\end{eqnarray}}
\newcommand{\ba}{\begin{array}}
\newcommand{\ea}{\end{array}}
\begin{document}

\vspace*{0.1cm}

\begin{flushright}  
Fermilab-PUB-18-561-T
\end{flushright} 

\title{\boldmath  Ultraheavy resonances at the LHC: beyond the QCD background  \\[0.3cm]}

\author{Bogdan A. Dobrescu, Robert M.~Harris and Joshua Isaacson  } 

\affiliation{%Theoretical Physics Department,
Fermi National Accelerator Laboratory, Batavia, Illinois 60510, USA   }

\date{October 22, 2018 \\[1.6cm] } 

\begin{abstract} \noindent \normalsize
We study the theory and some experimental hints of ultraheavy resonances at the LHC.
The  production of an ultraheavy narrow particle may have a 
 larger rate than the QCD background even when the final state includes only hadronic jets.
We consider two classes of models that lead to 4-jet signals. In the first class a 
diquark scalar decays into two vectorlike quarks. In the second one a coloron decays into two
color-octet scalars or into a pair of vectorlike quarks. 
We show that a diquark as heavy as 11.5 TeV, or a coloron as heavy as 8.5 TeV may be discovered at the LHC.
We point out that a CMS 4-jet event may be due to an 8 TeV resonance decaying into two secondary particles, 
each with a mass of 1.8 TeV. We find that the  QCD background  
with a 4-jet mass and dijet masses that equal or exceed those of the CMS event is approximately $5\times 10^{-5}$ events in 78 fb$^{-1}$ of data,
while the diquark signal  could have easily produced that event. 
The diquark also decays directly into two jets, which may be the origin of some of the  
 three other events of mass near 8 TeV observed by ATLAS and CMS.
 \end{abstract}

\maketitle
%%%%%%%%%%%%%%%%%%%%%%%%%%%%%%%%%%%%%%%%%%%%%%%%%
\newpage

%\vspace*{1cm}

%\bigskip \bigskip 

\hypersetup{linktocpage} % the table of content can have hyperlink
\tableofcontents
\hypersetup{linkcolor=purple}   %   blue} 

\makeatletter
\let\toc@pre\relax
\let\toc@post\relax
\makeatother

\bigskip \bigskip \bigskip 

%\newpage
%%%%%%%%%%%%%%%
\section{Introduction}
\label{sec:intro}
\setcounter{equation}{0}

The ATLAS and CMS experiments at the LHC have been pushing the limits of the Standard Model (SM) to higher and higher energies over the last eight years. 
New particles that can be produced in the $s$ channel have a large production cross section due to 
resonant enhancement, and yet may be narrow enough to be easily observed above a smooth background.
For example, the lower limit from dijet resonance searches (for reviews, see \cite{Han:2010rf,Harris:2011bh,Dobrescu:2013coa}) on the mass of a $Z'$ boson that couples to quarks the same way as the $Z$ boson has increased from 0.74 TeV before the start of the LHC \cite{Aaltonen:2008dn} 
to 1.7 TeV at the end of the 8 TeV run of the LHC \cite{Khachatryan:2015sja}, and is now 2.7 TeV \cite{CMS-PAS-EXO-17-026}. 
It is compelling to ask what is the highest-mass resonance that can be discovered or excluded
by the end of the upcoming high-luminosity run of the LHC.

Particles of mass above a few TeV that couple to two quarks, usually called diquarks, are 
expected to be produced with the highest cross section while still having a narrow width.
The reason is that the parton distribution functions (PDFs) of the up and down quarks are much larger than those of 
other proton constituents when the momentum carried by the parton is large.
While it is difficult to construct a renormalizable theory that includes  a spin-1 diquark (see {\it e.g.}, \cite{Assad:2017iib}),
spin-0 diquarks can easily be included in renormalizable extensions of the SM. 
In particular, color-sextet diquarks \cite{Mohapatra:2007af,  Han:2009ya, Berger:2010fy} that couple to $u\, u$ or $u\, d$ have the largest production cross section. 

Even though resonances produced from quark-antiquark initial states have smaller production rates,
they are still particularly interesting as they probe the existence of  gauge bosons beyond the SM. 
Among possible new gauge bosons, the coloron (a spin-1, color-octet particle)
has the special feature that it is associated with a non-Abelian gauge group which may be asymptotically free. 
As a result, the coloron coupling to quarks may be relatively large in a completely specified renormalizable
theory, especially if its interactions are flavor-independent \cite{Chivukula:1996yr}. By contrast, other gauge bosons, such as a $Z'$, cannot have a large coupling  
unless  the UV completion at a scale near the gauge boson mass  is some unidentified strongly coupled theory.
The coloron is associated with an extension of the QCD gauge group: $SU(3)_1 \times SU(3)_2$, which is spontaneously 
broken down to its diagonal subgroup, $SU(3)_c$,  responsible for the massless gluon. 
The simplest origin of the spontaneous symmetry breaking is a scalar field $\Sigma$  that transform as $(3,\bar 3)$ under the extended gauge group. The theory with the most general renormalizable potential for $\Sigma$ is referred to as 
the Renormalizable Coloron Model (ReCoM) \cite{Bai:2010dj,Chivukula:2013xka,Bai:2018jsr}. 

In this article we study theories with either a diquark scalar or a coloron of mass above half the collider center-of-mass energy, which can nevertheless be sufficiently weakly coupled to have a narrow width.
We show that such ultraheavy narrow resonances can be produced with a rate larger than that of the QCD background 
even when the final state consists only of jets.
After estimating the ultimate reach of the LHC in searches for narrow dijet resonances of these types, 
we consider the decay of the ultraheavy particle  in a pair of new TeV-scale particles. 
Specifically, we introduce a vectorlike quark which interacts in pairs with the diquark or the coloron, and decays predominantly into two jets. Thus, the signature is an ultraheavy 4-jet resonance, with each of two pairs of jets forming a dijet resonance
of mass equal to that of the vectorlike quark. 
A similar signature arises in the ReCoM from the coloron decay into scalars that are part of the  symmetry breaking field $\Sigma$. 
Related processes in strongly coupled theories have been studied in  \cite{Bai:2011mr,Kilic:2008pm}.

The CMS search for dijet resonances \cite{CMS-PAS-EXO-17-026}  has reported an unusual event, with four jets forming a resonance at 8.0 TeV. Those four jets
are grouped in two wide jets with equal masses, of 1.8 TeV.
We compute the probability that this event is due to QCD and find that it is approximately $5 \times 10^{-5}$.
We show that this event is consistent with a diquark decaying into a pair of vectorlike quarks, each then decaying into two jets. 
The diquark  also has a decay mode into two jets, which may  lead to an excess of dijet events of mass near 8 TeV.
The same CMS search reported a dijet event of mass at 7.9 TeV, while the latest ATLAS search for dijet resonances \cite{Aaboud:2017yvp} 
reported  two dijet events of mass in the 8.0--8.1 TeV range. Nevertheless,
these three dijet events may be due to a modest upward fluctuation of the QCD background.

We will not be concerned with underlying principles that may motivate the presence of the new particles discussed here.
We note though in passing that the color-sextet diquark scalars and vectorlike quarks 
are part of the 126 representation of the supersymmetric $SO(10)$ grand unification \cite{Mohapatra:1986uf}.
Theories that include a coloron and a vectorlike quark have been studied 
especially in conjunction with composite Higgs models. In fact, the term `coloron' has been introduced \cite{Hill:1991at} in the context of Higgs compositeness,
and later it was shown that natural models of this type must also include a vectorlike quark \cite{Dobrescu:1997nm}.

In Section \ref{sec:diquark} we present the diquark model and compute the signal and background 
cross sections. 
The coloron model and its signatures are analyzed in Section  \ref{sec:coloron}.
The events of mass near 8 TeV are discussed in Section \ref{sec:8TeV}.
Our conclusions are included in Section \ref{sec:conc}.

\newpage 

%\medskip

%%%%%%%%%%%%%%%%%%%%%%%%%%%%%%%%%%%%%%%%%%%%%%%%%%%%%%%%%%%%
\section{Diquark scalar plus vectorlike quark}
\label{sec:diquark}
\setcounter{equation}{0}

Let us consider a complex scalar field $S_{uu}$ which is a color-sextet of electric charge 4/3.
More precisely, $S_{uu}$ transforms under the SM gauge group 
$SU(3)_c\times SU(2)_W \times U(1)_Y$ as $(6, 1, +4/3)$, so that 
the only allowed renormalizable couplings of $S_{uu}$ to SM fermions are Yukawa interactions 
to right-handed up-type quarks. 

%%%%%%%%%%%%%%%%%%%%%%%%%%%%%%%%%%%%%%%
\subsection{Scalar interacting with two up quarks}
\label{sec:diquarkInt}

The interaction of the scalar diquark to two up quarks is required to produce  $S_{uu}$ at the LHC with a large rate.
We assume that a flavor symmetry is acting on right-handed 
up-type quarks such that  $S_{uu}$ couples only to the first generation:
\be
\dfrac{y_{uu} }{2} \; K^n_{ij} \; S_{uu}^n \;  \overline u_{R \, i} \,  u^c_{R \, j} \, + {\rm H.c.}  ~~~
\label{eq:diquarkYuk}
\ee
Here $y_{uu}$ is a dimensionless coupling. Without loss of generality we take $y_{uu} > 0$, because
any complex phase may be absorbed in a redefinition of the $u_R$ or $S_{uu}$ fields.
The factor of 1/2 is conventional for interactions that involve 
two fields of the same type. The upper script $c$  denotes the charge conjugated field.
The $i$ and $j$ indices on the quark fields label the triplet color states ($i,j = 1,2,3$),
and the upper index $n$ on the scalar field labels the sextet color states ($n=1,..., 6$).
The coefficients $K^n_{ij}$ are products of $SU(3)_c$ generators, which can be written as \cite{Luhn:2007yr}\cite{Han:2009ya}
\bear
&   \hspace*{-2cm}
 K^1_{ij} = \delta_{i1} \delta_{j1}      \;\;\; \;\;  ,    \;\;\; \;\; 
 K^2_{ij} = \dfrac{1}{\sqrt{2} } \left(  \delta_{i1} \delta_{j2} +  \delta_{i2} \delta_{j1}   \right)    \;\;\; \;\;  ,    \;\;\; \;\; 
 K^3_{ij} = \delta_{i2} \delta_{j2}  ~~,
\nonumber \\ [-2mm]
&
 \\ [-3mm]
&  \hspace*{-2cm}
K^4_{ij} = \dfrac{1}{\sqrt{2} } \left(  \delta_{i2} \delta_{j3} +  \delta_{i3} \delta_{j2}   \right)    \;\;\; ,    \;\;\; 
 K^5_{ij} = \delta_{i3} \delta_{j3}   \;\;\; ,    \;\;\; 
 K^6_{ij} = \dfrac{1}{\sqrt{2} } \left(  \delta_{i1} \delta_{j3} +  \delta_{i3} \delta_{j1}   \right)  ~~,
\nonumber
\eear
where $\delta_{ij}$ is the Kronecker symbol.

An example of flavor symmetry that prevents the coupling of the diquark to charm and top quarks
is a global $U(1)$ with both $u_R$ and $S_{uu}$ charged under it. The mass of the up quark then must arise from a higher-dimensional operator of the type $\phi_u H \bar Q_L u_R$, where $Q_L$ is a SM quark doublet, 
$H$ is the SM Higgs doublet, and $\phi_u$ is 
a scalar field which has a VEV and is charged under the global $U(1)$. That higher-dimensional operator 
provides an explanation for the smallness of the up quark mass. 
The flavor symmetry is convenient because it reduces the number of parameters as well as the number of diquark decay modes.
More important, it eliminates potentially large flavor-changing neutral  processes \cite{Fortes:2013dba}. 

The diquark interaction in Eq.~(\ref{eq:diquarkYuk}) leads to the $S_{uu} \to u u$ decay.
To leading order (LO), the width for this process is 
\be
 \Gamma (S_{uu} \to u \, u )  =   \dfrac{y_{uu}^2 }{32 \pi} \;  M_S  ~~.
 \label{eq:uuWidth}
\ee
If there are no other particles coupled to the diquark, then its  total  width, $\Gamma_S$, is given by
$ \Gamma (S_{uu} \to u \, u )$.
We are interested in a relatively narrow resonance, that produces a well defined peak in the invariant mass distribution. 
We will focus on a diquark with a total  width of up to 7\% of its mass. This allows the use of the narrow width approximation
in the computation of rates (for a recent study of the departures from the narrow width approximation see \cite{Chivukula:2017lyk}). Furthermore, this constraint on the width-to-mass ratio implies that the experimental resolution, about 4\% in CMS \cite{Sirunyan:2018xlo} for dijet resonances in the ultraheavy regime, is larger than the half-width.
Thus,  for comparison with the data, only the resolution determines the  mass range over which  the  predicted background should be integrated.
The  constraint  $\Gamma_S/M_{S} < 7\% $ translates to $y_{uu} < 2.7$.

\smallskip

%%%%%%%%%%%%%%%%%%%%%%%%%%%%%%%
\subsection{$S_{uu}$ diquark production and a dijet resonance at the LHC}
\label{sec:productionDiquark}

Next we analyze the production of the $S_{uu}$ scalar at the LHC.
The partonic cross section for $S_{uu}$ production at LO is given by
\be
\hat{\sigma} ( u u  \to S_{uu} ) = \dfrac{ \pi }{ 6  }  \,  y_{uu}^2 \;  \delta ( \hat{s} - M_{S}^2 )    ~~,
\ee
where $\sqrt{\hat s}$ is the center-of-mass energy of the partonic collision.
The cross section for $S_{uu}$ production in proton-proton collisions of energy $\sqrt{s}$ is 
\bear
\sigma \!\left( pp \to S_{uu} \right) = 
\dfrac{2}{s}  \int_0^1  \dfrac{dx}{x}  \int_0^{s x}  \! d \hat{s}  \;  u(x, M_{S}^2) \; u (\hat{s} / (s  x ), M_{S}^2 ) \;  \hat{\sigma} (uu \to S_{uu}  )  ~,
\eear
where $u(x,Q^2)$ is the PDF of the up quark  carrying momentum fraction $x$.
The LO production cross section is 
\be
\sigma \!\left( pp \to S_{uu}  \right) =  \dfrac{ \pi }{ 6 \, s }  \;  y_{uu}^2   \int_{M_{S}^2/s}^1 \dfrac{dx}{x}    \;
u\!\left( x, M_{S}^2 \right) \, u\!\left( M_{S}^2/ (s  x ) , M_{S}^2 \right) ~.
\label{eq:xsecDiquark}
\ee
Note that the production of  $S_{uu}^\dagger$, the antiparticle of  $S_{uu}$, leads to similar final states, but 
its cross section is much smaller because at partonic level it is due to $\bar u \bar u \to  S_{uu}^\dagger$, and the PDF of  $\bar u$ is much smaller than the one for the up quark.

\begin{figure}[t]
\hspace*{-3mm}\includegraphics[width=0.62\textwidth]{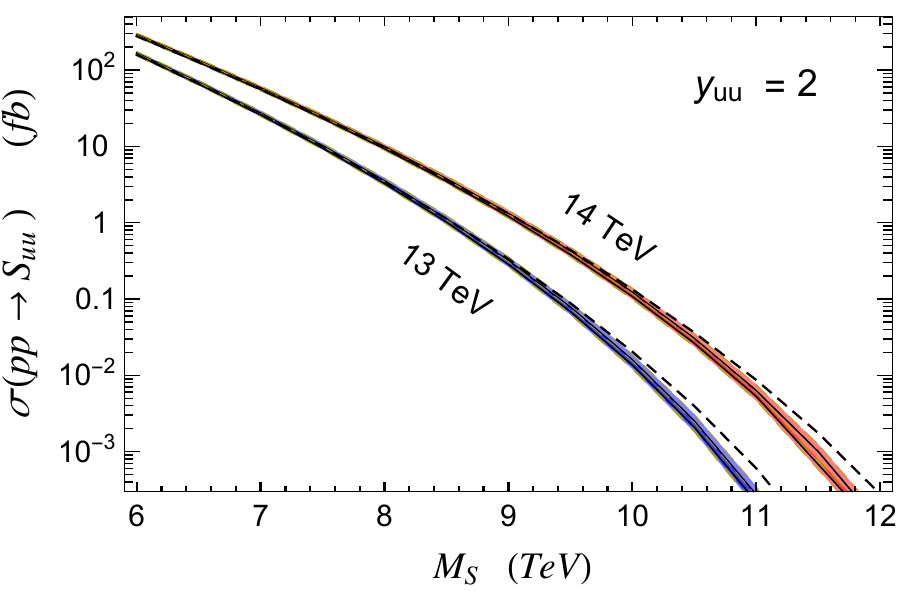}
\vspace*{-3mm}
\caption{Cross section for inclusive $S_{uu}$ diquark production at $\sqrt{s} = 13$ TeV (blue band) and 14 TeV (red band), computed at NLO
as a function of the $S_{uu}$ mass. The diquark coupling is fixed here at $y_{uu} = 2$, which corresponds to  a
width $\Gamma( S_{uu} \to uu) /M_S = 4\% $
(the cross section scales as $y_{uu}^2$).
The bands cover the central values obtained with the 
PDF sets CT14 \cite{Dulat:2015mca}
and MMHT2014 \cite{Harland-Lang:2014zoa} at NLO, as well as their $1\sigma$ uncertainties. The black dashed lines represent  the central values obtained with the NNPDF3.1 set \cite{Ball:2017nwa} at NLO.
 \\ [-3mm] }
\label{fig:xsecDiquark}
\end{figure}

The next-to-leading-order (NLO) QCD corrections to diquark production 
have been computed in \cite{Han:2009ya}, and yield an increase of about 16\% over the LO result.
Using the full analytic expression for the NLO cross section,  
we find the predicted $S_{uu}$ production cross section at the LHC shown in 
Figure~\ref{fig:xsecDiquark}: the blue band is for  $\sqrt{s} = 13$ TeV, while the 
red band is for  $\sqrt{s} = 14$ TeV.
 We used the current PDF sets at NLO from both 
CT14 \cite{Dulat:2015mca} and MMHT2014  \cite{Harland-Lang:2014zoa};
each band covers the values obtained with both sets as well as their uncertainties at the 68\% CL.
For comparison, the central values obtained with the PDF set NNPDF3.1 \cite{Ball:2017nwa} at NLO are shown 
as dashed black lines.  The cross sections  obtained with NNPDF3.1 become notably higher than the ones obtained with CT14 or MMHT2014 only for $M_S/\sqrt{s} \gtrsim 0.8$. 
The PDF values and uncertainties were obtained through the use of the  LHAPDF6 code~\cite{Buckley:2014ana}.

 The coupling of $S_{uu}$ to up quarks 
is fixed in Figure~\ref{fig:xsecDiquark} at $y_{uu} =2$, but the results shown there can be easily scaled by  $y_{uu}^2$
to obtain the cross section for any other value of the coupling.
Note that $y_{uu} =2$ corresponds to $\Gamma( S_{uu} \to uu) /M_S = 4\% $,
while the total width for $S_{uu}$  is 7\% of its mass if it also couples to a vectorlike quark as discussed 
in Section \ref{sec:signalDiquark}.

Let us estimate the ultimate reach of the LHC in searches for the $pp \to S_{uu} \to jj$ process,
where each $j$ is a jet obtained from the hadronization of an up quark. 
If there are no other new particles that $S_{uu}$ can decay into (such as the vectorlike quark of Section \ref{sec:signalDiquark}), 
then $S_{uu} \to j j $ has a branching fraction very close to 100\%.
The  QCD background to a dijet resonance search at the 14 TeV LHC has been estimated in \cite{Yu:2013wta}
for jets of $|\eta_j| < 2.5$ and a pseudorapidity separation of the two jets $|\Delta \eta_{jj}| <1.3$.
Taking a mass window of 8\% around the resonance mass, the background has a cross section of about 4.8 fb at 
a mass of 6 TeV, and decreases to about 0.13 fb at a mass of 8 TeV. 
The acceptance for a  scalar resonance  to pass the kinematic cuts mentioned above  is around 1/2.
The signal cross section   times the acceptance is  $30 y_{uu}^2$ fb for $M_S = 6$ TeV and $1.1 y_{uu}^2$  fb 
for $M_S = 8$ TeV. Thus, the signal-to-background ratio is already large (approximately 
$6.2y_{uu}^2$) at $M_S = 6$ TeV, and it grows with the resonance mass.

For $y_{uu}  \gtrsim  1$ and $M_S \gtrsim 8$ TeV, the background is so much smaller than the signal that the 
$|\Delta \eta_{jj}|$ cut may be removed. In that case the acceptance increases to about 0.9.
It is remarkable that for an $S_{uu}$ mass as large as 11.3 TeV, the production cross section 
at the 14 TeV LHC obtained with the CT14 set is approximately $2 \times 10^{-3}$ fb when $y_{uu} = 2$, which is large enough for 
about five dijet events to be observed with the anticipated 3000 fb$^{-1}$ of data.
For the  largest value of the coupling consistent with  $\Gamma_S /M_S < 7 \% $, namely  $y_{uu} = 2.7$,
an $S_{uu}$  diquark as heavy as 11.5 TeV may be discovered at the LHC.

\medskip
%\vspace*{-3mm}

%%%%%%%%%%%%%%%%%%%%%%%%%%%%%%%%%%%%%%%
\subsection{Up-type vectorlike quark and a  4-jet scalar resonance} 
\label{sec:signalDiquark}

Besides the $S_{uu}$ scalar, we now introduce another particle beyond the SM:  a vectorlike quark $\chi$, which has 
the same SM gauge charges as $u_R$, namely $(3, 1, +2/3)$. 
Given that the right- and  left-handed components of $\chi$ have the same gauge charges, there are 
two renormalizable interactions of $\chi$ with the diquark scalar, which are 
analogous to Eq.~(\ref{eq:diquarkYuk}):  
\be
\dfrac{1}{2} \; K^n_{ij} \; S_{uu}^n \;  \left(  y_{\chi_R} \, \overline \chi_{R \, i} \,  \chi^c_{R \, j} 
+ y_{\chi_L} e^{i\beta_\chi} \, \overline \chi_{L \, i} \,  \chi^c_{L \, j} \, \right) + {\rm H.c.}  ~~~
\label{eq:diquarkYukChi}
\ee
Since any complex phase of the coupling parameter $y_{\chi_R}$ may be absorbed in the $\chi_R$ field, we take $y_{\chi_R} \ge 0$.
The phase of the $\chi_L$ field is then fixed by the mass term $m_\chi \overline \chi_R\chi_L$, so the 
complex phase  of the second term in Eq.~(\ref{eq:diquarkYukChi}) cannot be removed. 
The coefficient of that term, $ y_{\chi_L} e^{i\beta_\chi}$, is the most general one 
with the dimensional parameters satisfying $y_{\chi_L} \ge 0$  and $0 \le \beta_\chi < 2\pi$.

The flavor symmetry introduced in Section~\ref{sec:diquarkInt} allows the interactions (\ref{eq:diquarkYukChi})
only if $\chi_R$ and $\chi_L$ have the same transformation property as $u_R$ under the global $U(1)$.
For simplicity, we assume that a coupling of $S_{uu}$ to a $\chi$ and an up quark has a negligibly small coefficient.
This can be a consequence of an approximate $Z_2$ symmetry, given by  invariance under the $\chi \to - \chi $ transformation.
As a result, a $\overline \chi_L u_R$ mass mixing is suppressed.

\begin{figure}[t]
\hspace*{-1mm} \includegraphics[width=0.493\textwidth]{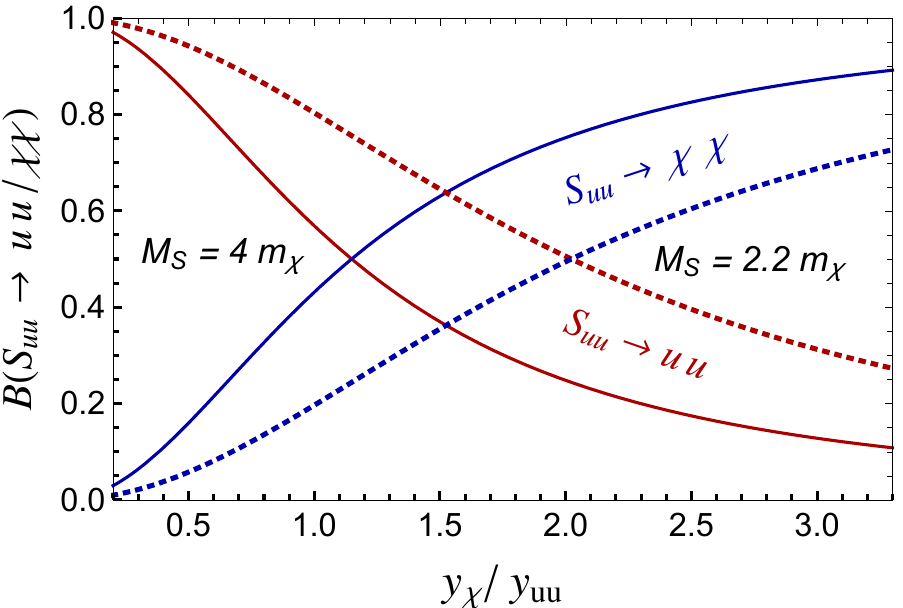} \hspace*{1mm}  
\includegraphics[width=0.48\textwidth]{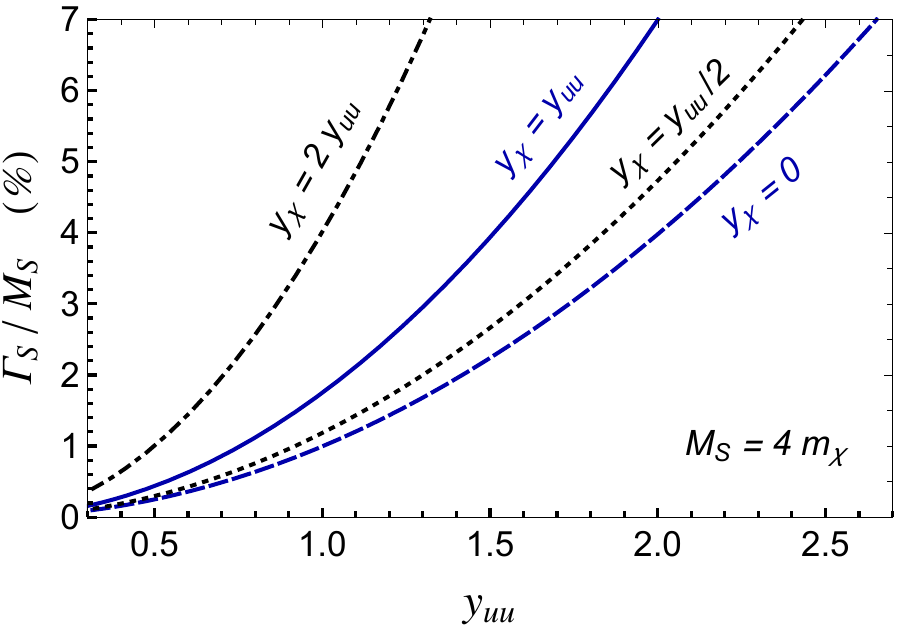} 
\vspace*{-11mm}
\caption{ {\it Left panel:} Branching fractions of the $S_{uu}$ diquark as a function of the combination of couplings
$y_\chi /y_{uu}\,$, where  $y_\chi \equiv \sqrt{y_{\chi_R}^2+y_{\chi_L}^2}$,
for a vectorlike quark mass $m_\chi = M_S/4$ (solid lines) or $M_S/2.2$ (dotted lines).
 {\it Right panel:} Width-to-mass ratio of $S_{uu}$ as a function of the up quark coupling $y_{uu}$,
 for $m_\chi = M_S/4$  and $y_\chi /y_{uu} = 2$, 1, 1/2 or 0.
 \\ [-1mm] }
\label{fig:BRdiquark}
\end{figure}

The mass of $\chi$ is chosen to satisfy $m_\chi < M_S/2$, so that the interactions Eq.~(\ref{eq:diquarkYukChi}) lead to 
a diquark decay into two vectorlike quarks. The LO partial width is given by
\be
\Gamma (S_{uu} \to \chi \, \chi)  =   \dfrac{1}{32 \pi} \left( y_{ \chi_R}^2  + y_{ \chi_L}^2  \right) \;  M_S  \;   \left( 1 - \dfrac{2 m_\chi^2 }{M_{S}^2 } \right)\left( 1 - \dfrac{4 m_\chi^2 }{M_{S}^2 } \right)^{\! 1/2}  ~~.
\label{eq:SchiWidth}
\ee
This partial width and the one given in Eq.~(\ref{eq:uuWidth}) give the branching fractions of $S_{uu} \to \chi \chi$ and $S_{uu} \to u u$  shown in the left panel of Figure~\ref{fig:BRdiquark} for two values of the mass ratio $M_S/m_\chi$. 
The width-to-mass ratio of $S_{uu}$ is shown in the right panel of Figure~\ref{fig:BRdiquark} for four values of  $\sqrt{y_{\chi_R}^2+y_{\chi_L}^2}/y_{uu}$.

The diquark interaction with the vectorlike quark in Eq.~(\ref{eq:diquarkYukChi}) is invariant under the $\chi \to - \chi $ transformation.
If this $Z_2$ symmetry were not broken by other Lagrangian terms, then the vectorlike quark would be stable.
We assume that the dominant $Z_2$-breaking interaction is a dimension-5 operator:
\be
\frac{C_g}{M_\star}  \, S_{uu}  \, \overline u_R \, \gamma^\mu \gamma^\nu \, T^a \chi_L  \, G^a_{\mu\nu}  + {\rm H.c.}  ~~,
\label{eq:dim5}
\ee
where $G^a_{\mu\nu}$ is the gluon field strength, $T^a$ are the $SU(3)_c$ generators,
$C_g$ is a dimensionless coefficient, and $M_\star$ is the mass of a very heavy particle which was integrated out. As this operator arises from a loop process, the $C_g$ coefficient is expected to be of the order of $1/(16 \pi^2)$ or smaller.
The only 2-body decay induced by the dimension-5 operator is $\chi \to u g$, which leads to a dijet resonance of mass $m_\chi$.
Another mechanism for a decay of a vectorlike quark into a gluon and a SM quark is discussed in \cite{Kim:2018mks}.

If there is some mass mixing between $\chi$ and the $u_R$ quark, then $\chi$ may also decay into $W d$, $Z u$ or $h^0 u$, where  $h^0$ is the SM Higgs boson. If that mixing is negligible, then various exotic decays of $\chi$ 
may have large  branching fractions \cite{Dobrescu:2016pda}.  We will ignore such processes here, so that 
$\chi$ decays into two jets with a branching fraction of close to  100\%. 
Note though that whatever loops  induce operator (\ref{eq:dim5}), there must be 
similar loop processes with the gluon replaced by a photon or a $Z$; the rates of these
processes are model dependent, but typically the gluon process has a rate larger by a factor of $3\alpha_s/\alpha \approx 40$.
The total width of $\chi$ is very small, below $10^{-5}$ of its mass; nevertheless, the $\chi$ decays are prompt for a large range of parameters.

\begin{figure}[t]
\hspace*{-1mm} 
 \includegraphics[width=0.42\textwidth]{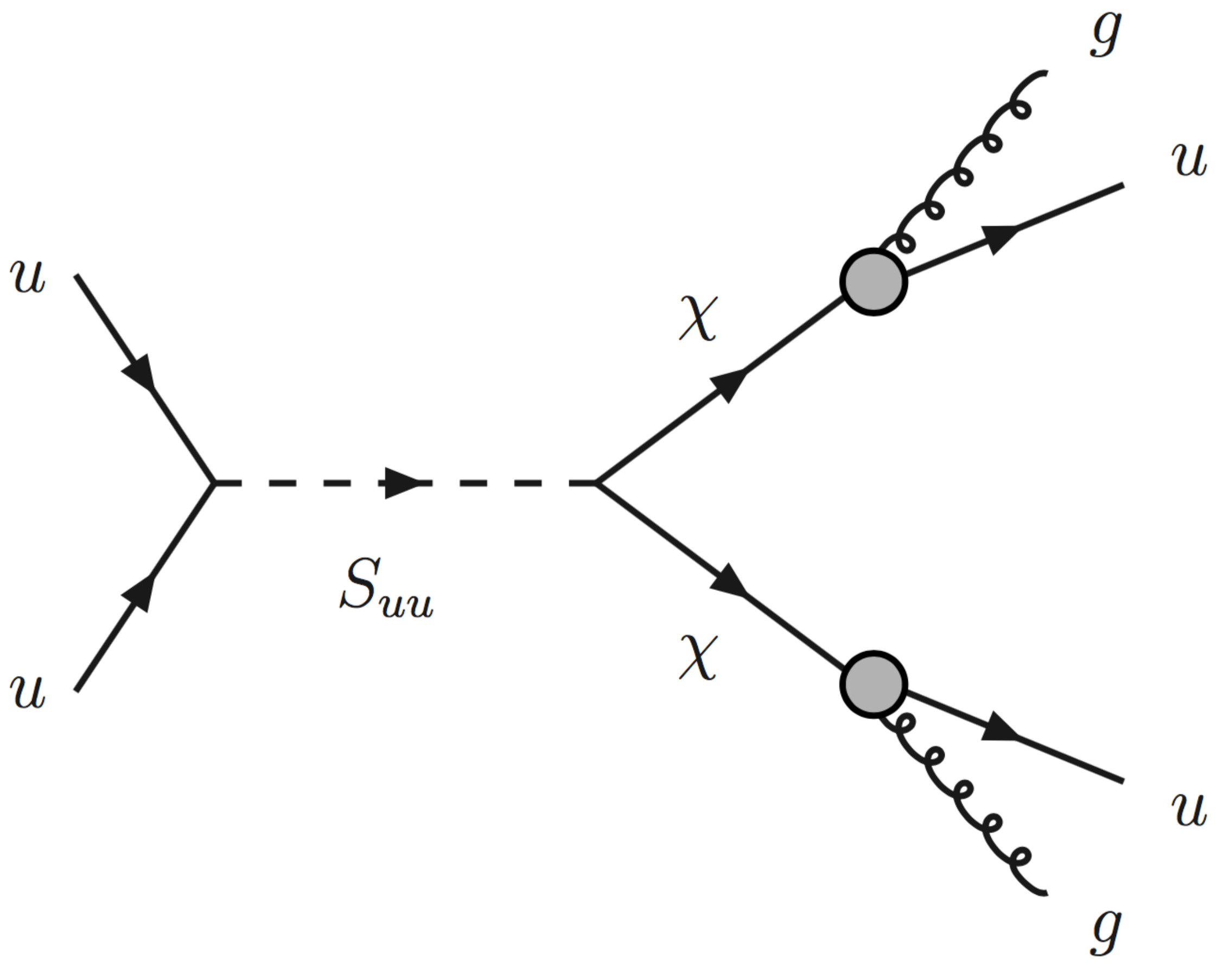}
\vspace*{-5mm}
\caption{Diquark production followed by decay into a pair of  vectorlike quarks, each of them then decaying at one loop into 
a gluon and an up quark. 
 \\ [-3mm] }
\label{fig:diagramsDiquark}
\end{figure}

The $S_{uu} \to \chi \chi \to (g \, u )(g \, u )$ process (see Figure \ref{fig:diagramsDiquark})  leads to resonant production of 
a pair of  equal-mass dijet resonances.
For $M_S \gg m_\chi$, the signal is a pair of wide jets, each with 2-prong substructure.
The rate for this process is given by the cross section of Figure~\ref{fig:xsecDiquark} multiplied by the branching fraction 
of Figure~\ref{fig:BRdiquark}. Note that the 4-jet resonant signal and the ultraheavy dijet resonance signal 
are both large for a range of parameters.

There is an additional new-physics contribution to the final state with a pair of dijet resonances: 
QCD production of $\chi \bar\chi $ leads to nonresonant production of  $(g u )(g \bar u) $. 
This nonresonant process has a larger cross section than the diquark induced one 
when $m_\chi \ll M_S$. However, the former process does not lead to a resonance in the invariant mass distribution 
of the four jets, and also it does not typically produce wide jets with substructure. Thus, the contribution of the nonresonant 
4-jet process to the signal of an ultraheavy diquark is negligible.  Nevertheless, the nonresonant $\chi \bar\chi$ production
is constrained by searches for a pair of dijet resonances, leading to a lower limit on the vectorlike quark  mass. 
The NLO cross section for $p p \to \chi \bar\chi  $ \cite{Yu:2013wta}
at the 13 TeV LHC is approximately 120 fb for $m_\chi = 0.8$ TeV,
and 82 fb for $m_\chi = 0.9$ TeV.
The CMS search of this type with 36 fb$^{-1}$ of data at $\sqrt{s} = 13$ TeV 
sets a limit $m_\chi \gtrsim 0.9$ TeV (see left panel of Fig.~11 in  \cite{Sirunyan:2018rlj}),
while the ATLAS search \cite{Aaboud:2017nmi} with a similar amount of data sets a slighly looser limit.

The $S_{uu}$ signal has a large acceptance, ${\cal A}_{4j} \approx 0.8$, in 
a search for an ultraheavy 4-jet resonance, in which it is required that two dijet resonances have approximately equal mass.
For a benchmark in the parameter space where 
 $m_\chi = M_S/4$ and $\sqrt{y_{\chi_R}^2+y_{\chi_L}^2} =  y_{uu}= 2$,
 the width-to-mass ratio is 7\% and  the branching fraction $B(S_{uu} \to \chi\chi) = 43\%$;  
the cross section times ${\cal A}_{4j} B(S_{uu} \to \chi\chi)$ is then 
$3.6 \times 10^{-2}$ fb at $M_S = 10$ TeV,  and 
$1.5 \times 10^{-3}$ fb at $M_S = 11$ TeV.
To estimate the reach in the 4-jet channel, we next compute the background to this process. 

\vspace*{-1mm}

%%%%%%%%%%%%%%%%%%%%%%%%%%%%%%
\subsection{QCD 4-jet background} 
\label{sec:backDiquark}

\begin{figure}[t]
\hspace*{-11mm} 
 \includegraphics[width=0.33\textwidth]{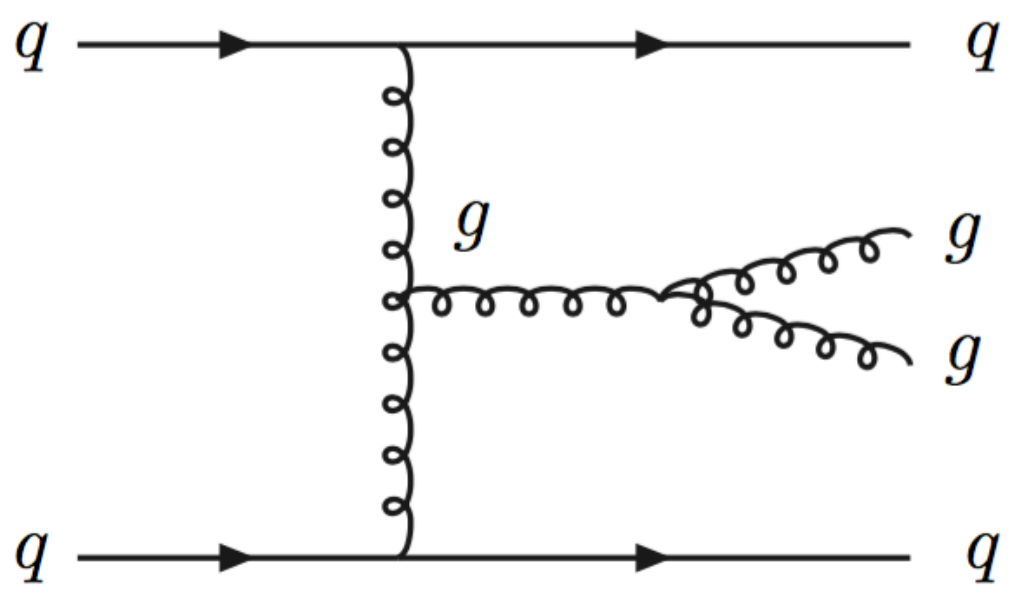}  \hspace*{18mm}   \includegraphics[width=0.27\textwidth]{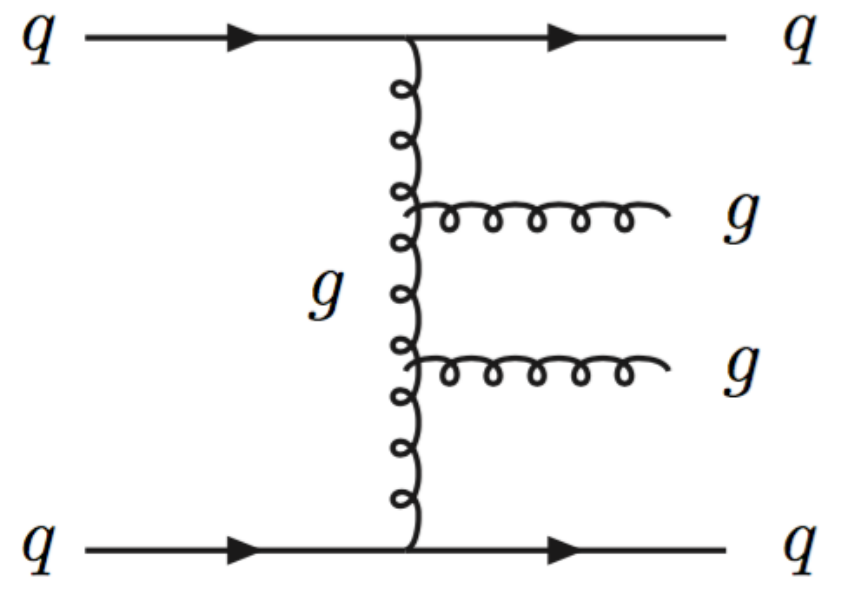}
\\ [5mm]
  \includegraphics[width=0.33\textwidth]{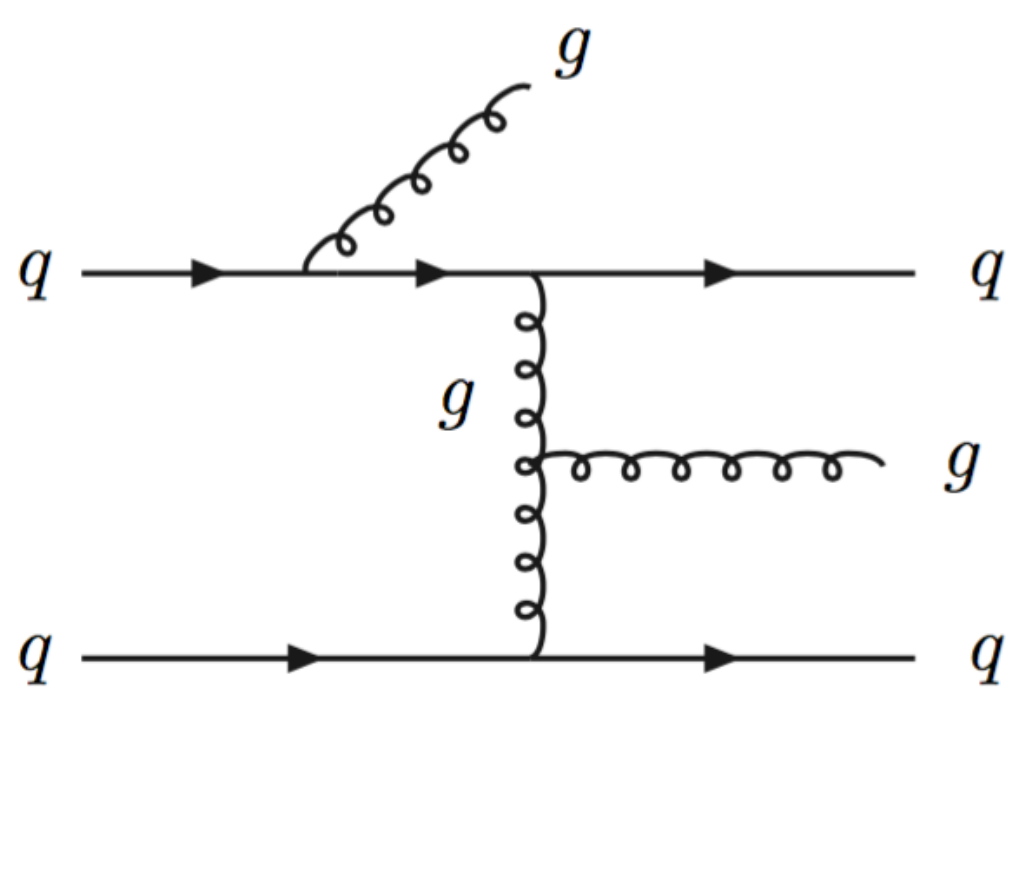} \hspace*{15mm}   \hspace*{8mm}     \includegraphics[width=0.36\textwidth]{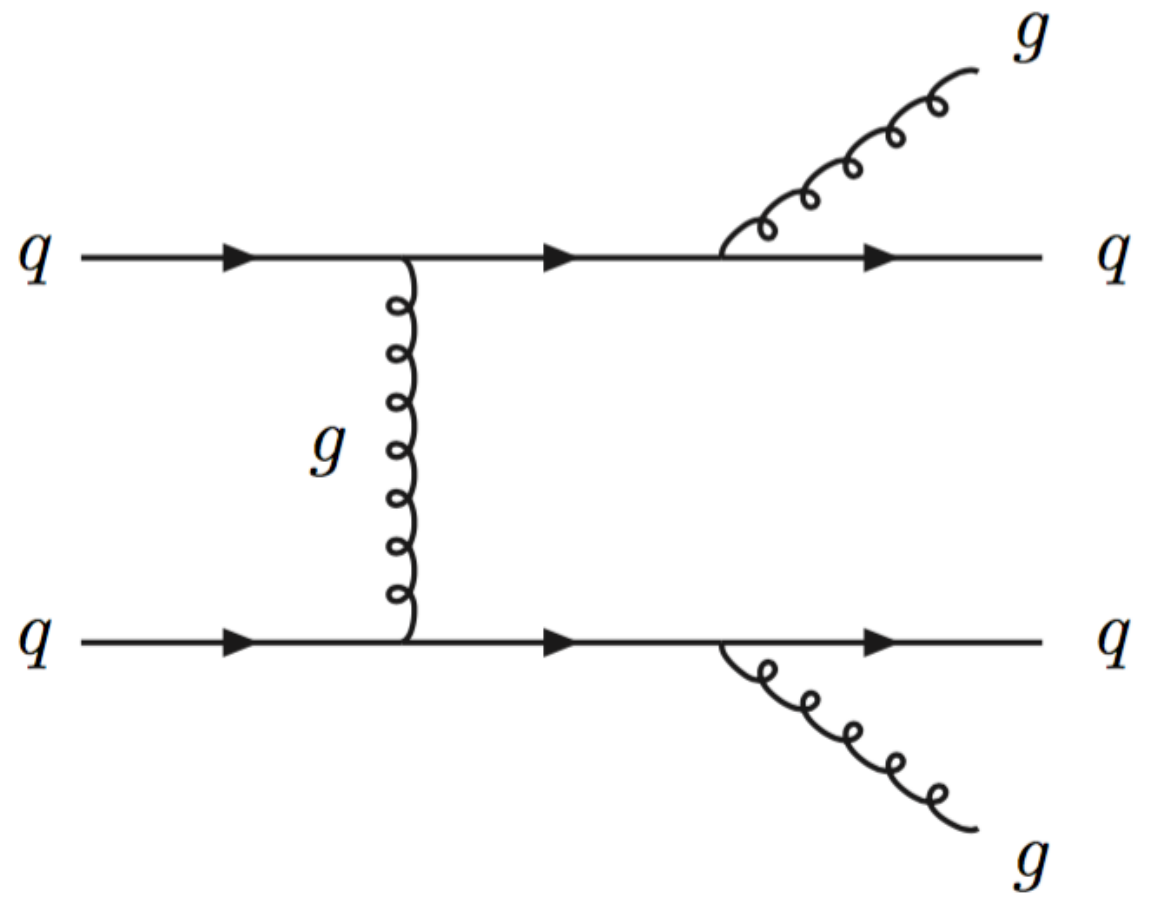}   \\ [1mm]
\includegraphics[width=0.4\textwidth]{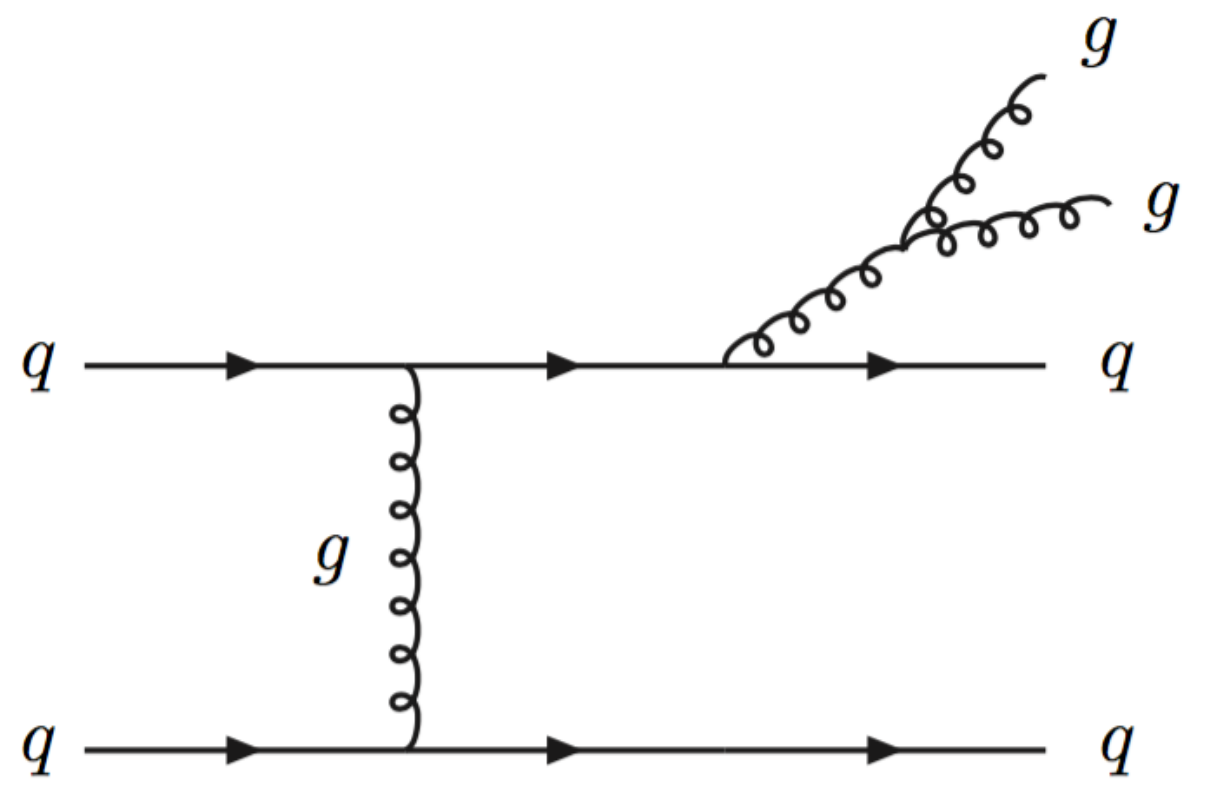}   \hspace*{12mm}   \includegraphics[width=0.36\textwidth]{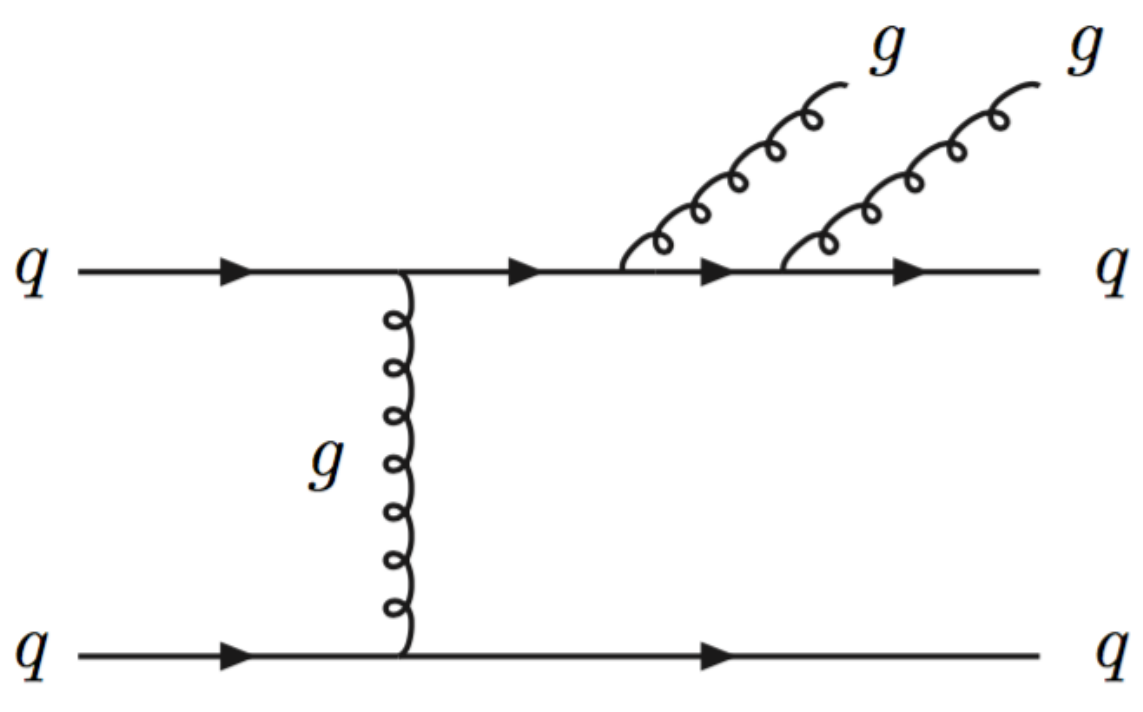}
%\vspace*{-1mm}
\caption{Representative diagrams for the dominant QCD background to the $4j$ signal, namely quark-quark collisions with two
radiated gluons. Eight other diagrams (not taking into account crossing symmetries) similar with the third and fourth diagrams, 
and six other diagrams similar with the last two are not shown.
\\ [-2mm] }
\label{fig:QCDdiagrams}
\end{figure}

The main background to the   $pp \to S_{uu} \to \chi \chi \to (g \, u )(g \, u )$ signal is QCD production of four jets. For the very large masses of the 4-jet resonance considered here,
the dominant production arises from the quark-quark initial states. Furthermore, the final states with two quarks and two gluons 
have larger contributions than those with three quarks and one anti-quark. Altogether there are twenty Feynman diagrams (ignoring crossing symmetries) contributing to this process,
with some representative ones shown in Figure \ref{fig:QCDdiagrams}.

\newpage

We have computed the cross section for the QCD $pp \to 4j$ process using MadGraph \cite{Alwall:2014hca} at LO 
(more precisely MadGraph5\_aMC@NLO v2.6) with the 
% \newpage \noindent
following kinematic cuts imposed using MadAnalysis 5 \cite{Conte:2012fm}:  \\  % [1mm] 
%\noindent\noindent
{\tiny \ \  \hspace*{6.6mm}  $\bullet$ \ } {\rm transverse momentum of each jet}: $p_{Tj} > 400$ GeV;  \\
{\tiny \ \  \hspace*{6.6mm} $\bullet$ \ } {\rm pseudorapidity of each jet}: $|\eta_j| < 2.5$;   \\
{\tiny \ \  \hspace*{6.6mm} $\bullet$ \ } {\rm separation for each pair of jets}:  $\Delta R_{jj'} > 0.4$;  \\
{\tiny \ \  \hspace*{6.6mm}  $\bullet$ \ } {\rm two pairs of jets, with the pair masses satisfying $m_{jj1}, m_{jj2} > 1 $ TeV
as well as $(m_{jj1} - m_{jj2})/ \overline{m}_{jj} < 7\%$, where  $\overline{m}_{jj} = (m_{jj1}+m_{jj2})/2$.

In Figure \ref{fig:QCDxsec} we show the QCD 
cross section at LO in mass bins of 100 GeV as a function of the 4-jet invariant mass, 
obtained with the NNPDF2.3~\cite{Ball:2017nwa} PDF set at LO. 
The result with the CT14  set \cite{Dulat:2015mca} at LO is higher by up to 20\%, and the 
result with the  MMHT2014 set \cite{Harland-Lang:2014zoa}  at LO is lower by about 15\%.
To obtain the background in a mass window equal to  8\% of the diquark mass,
we sum over a number of bins given by $0.8 M_S/(1$ TeV). This background is smaller  by at least  three orders of magnitude 
 than the diquark production cross section shown in Figure~\ref{fig:xsecDiquark}. 

\begin{figure}[t]
\hspace*{-3mm}\includegraphics[width=0.61\textwidth]{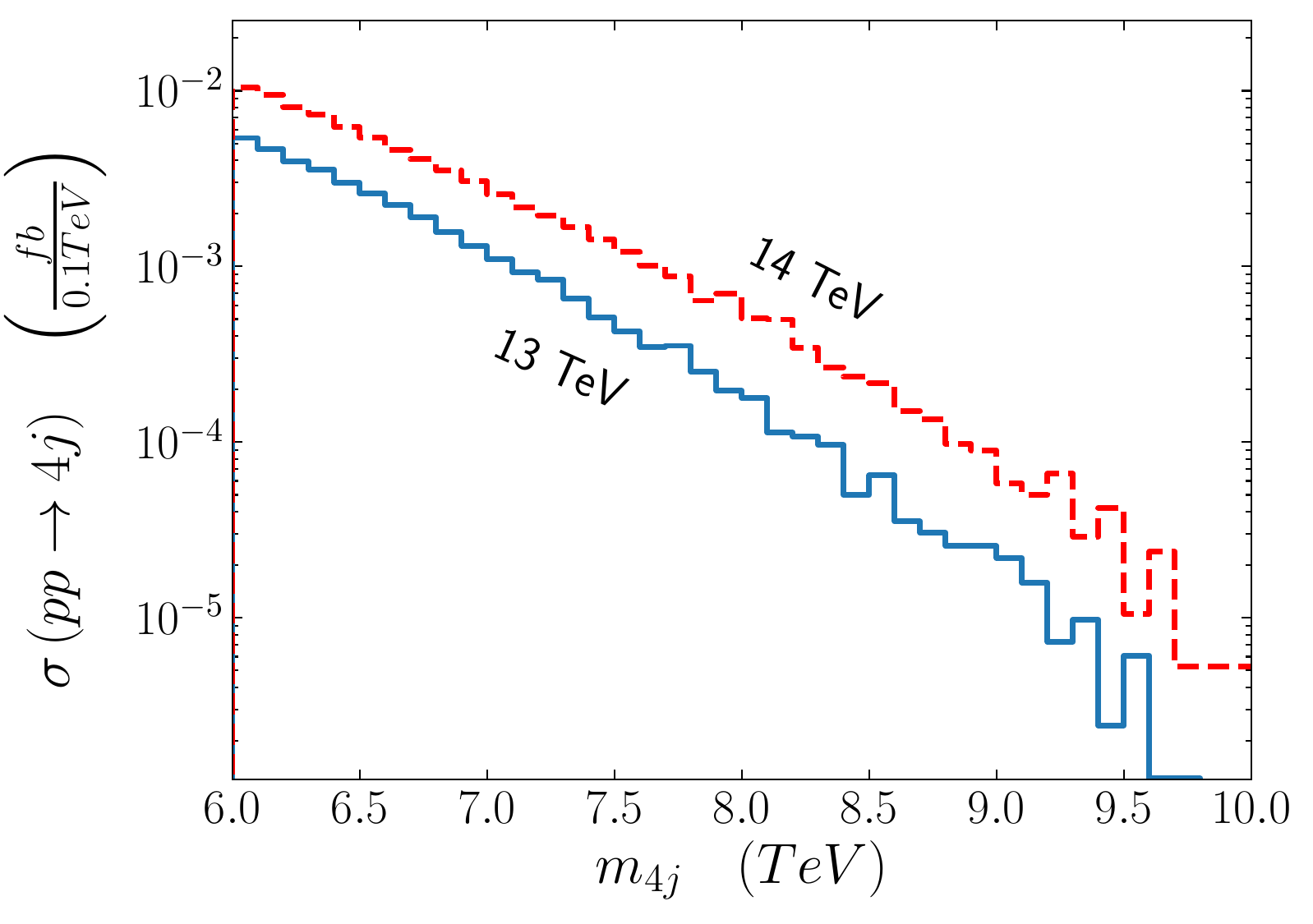}    % .eps}
\vspace*{-3mm}

%\vspace*{-1mm}
\caption{Cross section for QCD  4-jet production at the LHC at $\sqrt{s} = 13$ TeV and 14 TeV, 
in mass bins of 100 GeV, as a function of the $4j$ mass. 
The cuts imposed on the leading four jets are: $p_{Tj} > 400$ GeV, $|\eta_j| < 2.5$, $\Delta R_{jj'} > 0.4$.  
In addition, two pairs of jets are each required to have a mass 
above 1 TeV, and  the relative mass difference of the two 
pairs must be below 7\%.
 \\ [-2mm] }
\label{fig:QCDxsec}
\end{figure}

An $S_{uu}$ diquark of mass  at 11 TeV (see the benchmark in Section \ref{sec:signalDiquark})  may produce  four or  five 4-jet events 
with 3000 fb$^{-1}$ of integrated luminosity. Given that the background is below $10^{-2}$ events, we conclude that a narrow particle as heavy as 
11 TeV may be discovered in the 4-jet channel by the end of the high luminosity run of the LHC. 
This mass reach is lower than that  computed in Section \ref{sec:productionDiquark} for the dijet channel  due to the smaller value of $y_{uu}$ for fixed width-to-mass ratio,
and also the smaller branching fraction.

\medskip
%%%%%%%%%%%%%%%%%%%%%%%%%%%%%%%%%%%%%%%%%%%%%%%%%%%%%%%%%%%%
%%%%%%%%%%%%%%%%%%%%%%%%%%%%%%%%%%%%%%%%%%%%%%%%%%%%%%%%%%%%
\section{Coloron plus scalar or vectorlike quark}
\label{sec:coloron}
\setcounter{equation}{0}

We now turn to a different theory that includes LHC signals similar to those discussed in the previous section.
The ReCoM theory \cite{Bai:2010dj, Bai:2018jsr} is a gauge extension of QCD to $SU(3)_1 \times SU(3)_2$, with this extended gauge group 
spontaneously broken by a scalar field $\Sigma$ which transforms as $(3,\bar 3)$. 
For a large range of parameters, $\Sigma$ acquires a VEV that gives the coloron a mass while leaving the gluon massless \cite{Bai:2017zhj}.
All SM quarks transform 
as triplets under  $SU(3)_1$. Below the symmetry breaking scale, the theory consists of the usual QCD with the unbroken gauge group $SU(3)_c$ associated with a massless gluon, plus a massive spin-1 color-octet particle 
(the coloron) and some scalars particles discussed later on (see Section \ref{sec:signalColoron}).

The coloron ($G'$) couples to the SM quarks as follows:
\be
g_s \tan\theta  \; G^{\prime a}_\mu \,  \overline q \, \gamma^\mu \, T^a \, q   ~~,
\label{eq:coloronQuarks}
\ee
where %$T^a$ are the  $SU(3)_c$ generators, 
$g_s$ is the QCD gauge coupling (related to the strong coupling constant by 
$\alpha_s = g_s^2/(4 \pi)$), and $\tan\theta$ is the ratio of the $SU(3)_1 \times SU(3)_2$ gauge couplings. 
The range of $1/5 \lesssim \tan\theta \lesssim 5$ is the one where the theory is perturbative. For  $\tan\theta \gtrsim 5$ the coloron coupling to quarks becomes 
nonperturbative, while for $\tan\theta \lesssim 1/5$ the coloron self-coupling becomes nonperturbative.
We focus in this Section on the signals of an ultraheavy coloron.

%%%%%%%%%%%%%%%%%%%%%%%%%%%%%%%
\subsection{Coloron production}
\label{sec:production}

The partial width into SM quark-antiquark pairs, summed over the six SM flavors, is given at LO by
\be
\sum_q \Gamma (G' \to q\bar q) = \alpha_s (M_{G'}) \, \tan^2\! \theta   \; M_{G'}  ~~~.
\label{eq:qqbarWidth}
\ee
We have neglected here the top quark mass, because the corrections are 
of order $(m_t /M_{G'})^2$, which is below $10^{-3}$ for  a coloron heavier than 6 TeV.
The strong coupling constant $\alpha_s (M_{G'})$ at a scale $M_{G'} \approx 6$ TeV is approximately 0.076 if there are no new
colored particles lighter than the coloron. Eq.~(\ref{eq:qqbarWidth}) implies that 
the ratio of total width to mass for the coloron satisfies $\Gamma_{G'} / M_{G'}   \ge  \alpha_s \tan^2\! \theta$. 
As mentioned in Section \ref{sec:diquarkInt}, we are interested in a narrow resonance, that produces a well-defined mass peak.
Imposing $\Gamma_{G'} / M_{G'} < 7\% $ sets un upper limit $\tan\theta \lesssim 1$.

The cross section for coloron production is 
\bear
\sigma \!\left( pp \to G' \right) =  
\dfrac{2}{s}  \int_0^1  \dfrac{dx}{x}  \int_0^{s x}  \! d \hat{s}  \, \sum_q  q(x, M_{G'}^2) \, \bar q (\hat{s} / (s  x ), M_{G'}^2) \;  \hat{\sigma} (q \bar q \to G' )  ~,
\eear
where $q(x,Q^2)$ is the PDF of quark $q=d,u,s,c,b$ carrying momentum fraction $x$, and
$\hat{s}$ is the center-of mass  energy of the partonic collision.
The production  cross section at the partonic level is given by \cite{Bai:2010dj}:
\be
\hat{\sigma} (q \bar q \to G' ) = \dfrac{ 16 \pi^2 }{ 9  }  \,  \alpha_s \tan^2 \!\theta \; \,  \delta ( \hat{s} - M_{G'}^2 )    ~.
\ee
The LO production cross section is 
\be
\sigma \!\left( pp \to G' \right) =  \dfrac{ 16 \pi^2 }{ 9 s }  \,  \alpha_s \tan^2 \!\theta   \int_{M_{G'}^2/s}^1 \dfrac{dx}{x}   \sum_q  \left( 
q  \left( x, M_{G'}^2 \right) \, \bar q   \left( M_{G'}^2/ (s  x ) , M_{G'}^2 \right) +  q \leftrightarrow \bar q 
 \rule{0mm}{4mm} \right) ~.
\label{eq:xsecColoron}
\ee
The uncertainties in the PDFs are larger in the case of an ultraheavy coloron than those discussed in Section \ref{sec:productionDiquark}. The reason is that the anti-quark PDFs  are less constrained at large $x$.
In the case of the NNPDF3.1  \cite{Ball:2017nwa} set 
the central value of the predicted cross section is larger than 
the CT14 \cite{Dulat:2015mca} one by a factor of about 13 over the mass range 6.5 TeV $\lesssim M_{G'} \lesssim 9.5$ TeV.
It appears  more reliable to use the CT14 or MMHT2014   \cite{Harland-Lang:2014zoa}  PDF sets, 
which employ a physical parametrization, than the NNPDF3.1 set  which is based on neural network fits to the data 
at significantly smaller $x$. 
We also note that the results obtained with  the
NNPDF sets at NLO or next-to-next-to-leading-order (NNLO) PDF sets are unphysical in some cases: 
 the cross section in Eq.~(\ref{eq:xsecColoron}) is negative for the central values of those sets.

Using Eq.~(\ref{eq:xsecColoron}) and the ManeParse package \cite{Clark:2016jgm} for reading the PDF files, 
we find the predicted coloron production cross section at the LHC  shown in 
Figure~\ref{fig:xsec} for the LO CT14 and MMHT2014 PDF sets, including the 
$1\sigma$-band of the MMHT2014 uncertainties.
The NLO corrections to coloron production have been computed in \cite{Chivukula:2011ng, Chivukula:2013xla}. For $M_{G'}$ 
above 6 TeV, the LO cross section is enhanced by a factor $K \approx 1.2$. However, the NLO PDF sets at large $x$ lead to a cross section  smaller by a factor of 2. 
Since the NLO PDFs are obtained from fits at much lower energies than those of the ultraheavy regime, we will use only 
the LO coloron cross section in what follows. 

\begin{figure}[t]
\hspace*{-1mm}\includegraphics[width=0.483\textwidth]{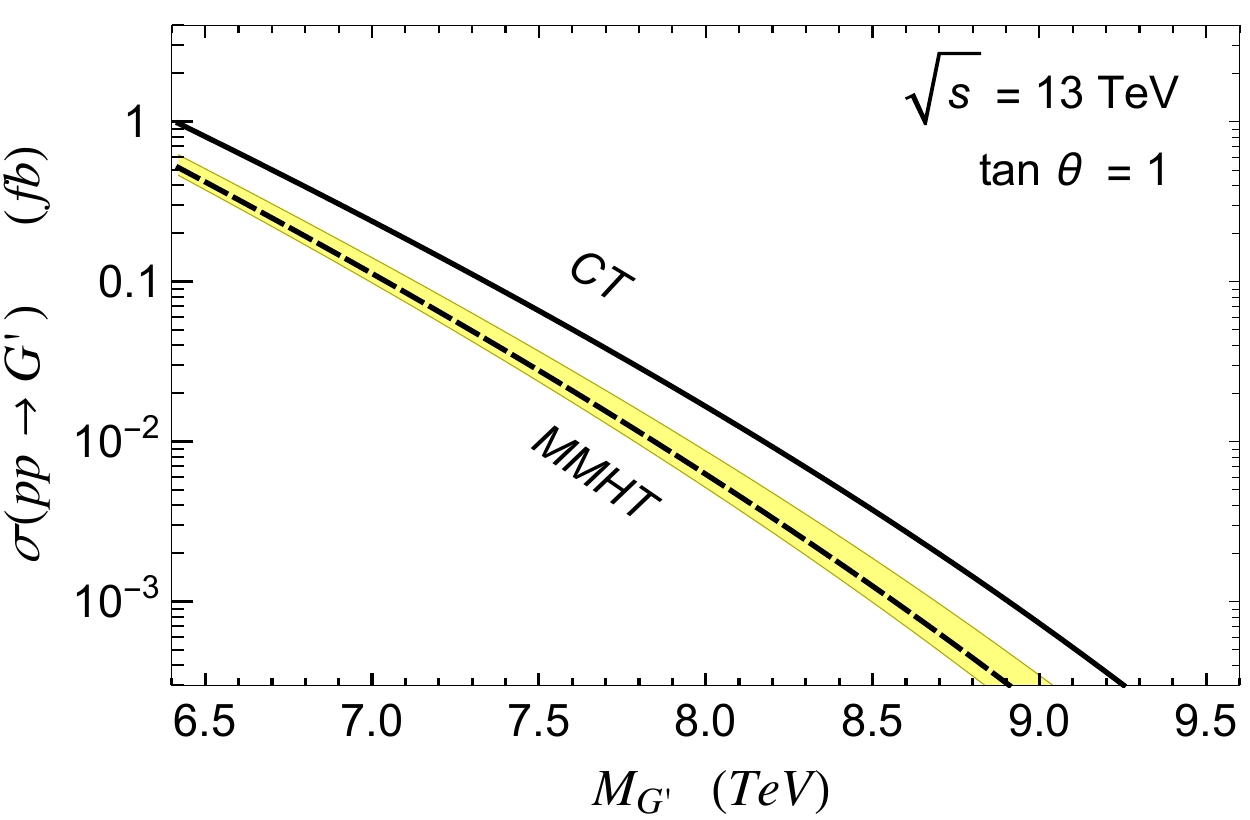} \hspace*{4mm}  \includegraphics[width=0.483\textwidth]{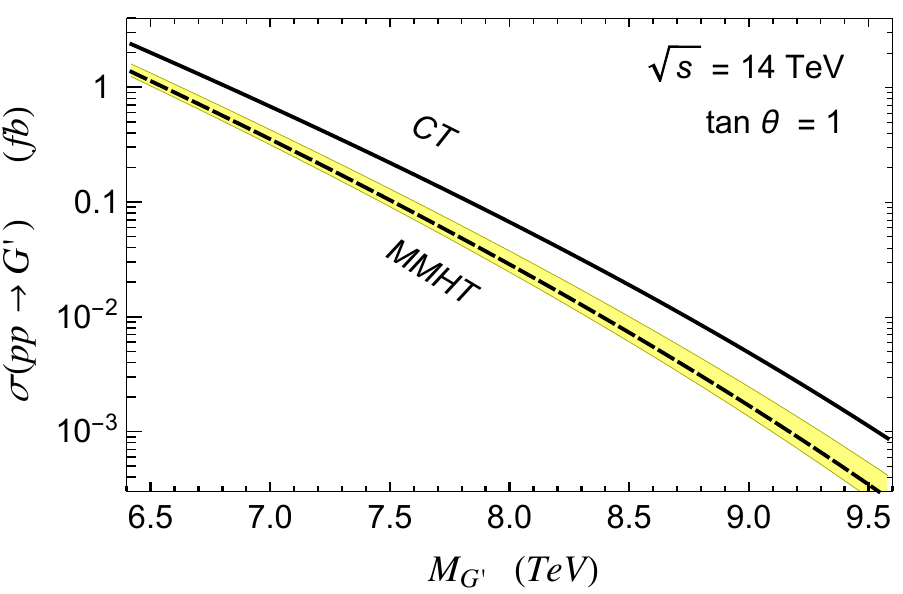}
\caption{LO coloron production cross section at $\sqrt{s} = 13$ TeV (left panel) and 14 TeV (right panel), 
as a function of the coloron mass, for the coloron coupling to quarks equal to that of the gluon ($\tan\theta = 1$). 
The LO PDF sets used here are CT14 \cite{Dulat:2015mca} (solid line) 
and MMHT2014 \cite{Harland-Lang:2014zoa} (dashed line is the central value, shaded band is the uncertainty).
The cross section scales as $\tan^2\! \theta$.
 \\ [-2mm] }
\label{fig:xsec}
\end{figure}

The acceptance for a high-mass vector resonance signal in dijet searches is around 0.4.
Using the background estimate discussed at the end of Section \ref{sec:productionDiquark},
we conclude from Figure~\ref{fig:xsec}  that for $\tan\theta = 1$ a coloron of mass up to 8 TeV 
can be discovered as a narrow dijet resonance with 3000 fb$^{-1}$ of data at 14 TeV.  
If the coloron can decay into new particles, then the mass reach of the LHC may be higher, as we discuss next.

\smallskip
%%%%%%%%%%%%%%%%%%%%%%%%%%%%%%
\subsection{Coloron signals with resonant pair of dijets}
\label{sec:signalColoron}

The  scalar field  responsible for breaking the extended gauge symmetry, $\Sigma$, includes the longitudinal degrees of freedom of the coloron, as well as three physical spin-0 particles \cite{Bai:2010dj, Bai:2018jsr, Chivukula:2013xka, Bai:2018wnt}: a color-octet $\Theta$, a
gauge-singlet CP-odd scalar $\phi_I$, and a gauge-singlet CP-even scalar $\phi_R$. 
For simplicity, we will assume in what follows that the gauge-singlet scalars are sufficiently heavy so that their production 
at the LHC can be neglected.
We refer to this particular case of the ReCoM as the ``coloron+scalar"  model.

We also consider an extension of the ReCoM that includes 
a vectorlike quark, $\chi$, which transforms under $SU(3)_1 \times SU(3)_2$ as $(1, 3)$, while the $\Theta$ scalar 
in this case is assumed heavier than half the coloron mass. We refer to this as the 
``coloron+quark"  model. There is some freedom in choosing the charges of $\chi$ under the SM gauge group.
For definiteness, we will assume that $\chi$ is an $SU(2)_W$ singlet, and has electric charge +2/3.

The coupling involving two $\Theta$ scalars and one coloron is proportional to the totally-antisymmetric color tensor ($f^{abc} $):
\be
- \dfrac{g_s}{2} \,  \left( \cot\theta - \tan\theta \right)   \,f^{abc}\, G_\mu^{\prime\,a} \,  \Theta^b  \,  \partial^\mu\,\Theta^c  ~~.
\label{eq:coloronScalars}
\ee
In the coloron+scalar model, this leads  to coloron decays into a pair of $\Theta$ scalars with a
partial width  \cite{Bai:2018jsr}
\be
 \Gamma (G' \to \Theta \Theta)  = \dfrac{1}{32}  \left( \cot^2\! \theta - 1 \right)^2 \; \left( 1 - \dfrac{4 M_\Theta^2}{M_{G'}^2 } \right)^{\! 3/2} \,  \sum_q \Gamma (G' \to q\bar q)  ~~,
 \label{eq:Theta2}
\ee
where we used the partial width into quarks given in Eq.~(\ref{eq:qqbarWidth}).
For $\tan\theta = 1$ this partial width vanishes, as a consequence of the $Z_2$ symmetry which interchanges the two $SU(3)$ gauge groups. 
As a result, the total width-to-mass ratio does not surpass 7\% for $\tan\theta < 0.95$ (see Figure \ref{fig:GpWidth}).

\begin{figure}
\hspace*{-3mm}\includegraphics[width=0.5\textwidth]{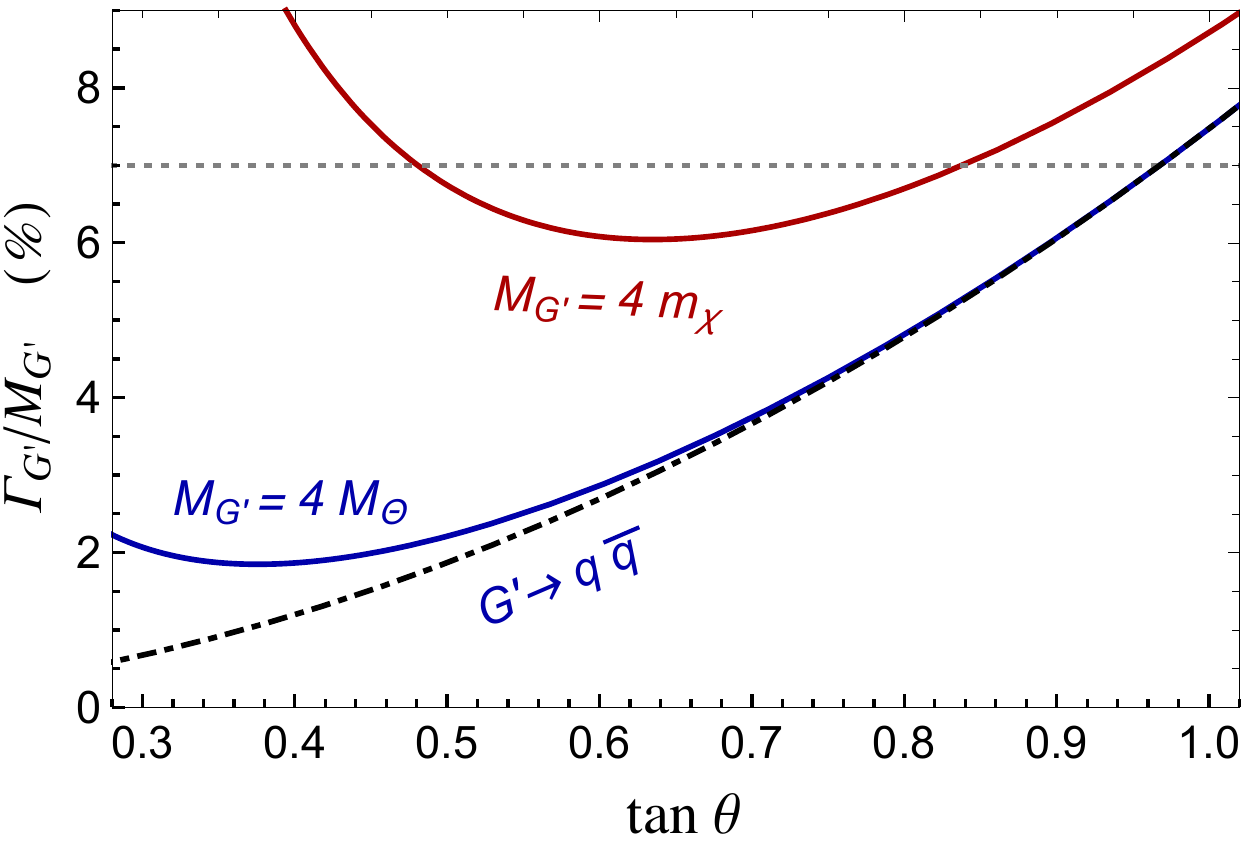}\vspace*{-1mm}
\caption{Ratio of total width to mass for the coloron ($G'$) in the models with color-octet scalar $\Theta$ (blue solid line)  or  vectorlike quark $\chi$ (red solid line) of mass $M_{G'}/4$,
as a function of the coupling normalization $\tan\theta$. The black dash-dotted line represents the $\Gamma_{G'}/M_{G'}$ ratio when only the decays into SM quarks are kinematically open. The coloron is considered a narrow resonance
when its width is below 7\% of the mass (marked by the dotted gray line).
 \\ [-1mm] }
\label{fig:GpWidth}
\end{figure}

In the coloron+quark model, the vectorlike quark $\chi$ couples to the coloron proportionally to $1/\tan\theta$:
\be
\dfrac{g_s}{ \tan\theta}  \; G^{\prime a}_\mu \,  \overline \chi \, \gamma^\mu \, T^a \, \chi   ~~.
\ee
Note that this is different than Eq.~(\ref{eq:coloronQuarks}) because $\chi$ is a triplet under the second $SU(3)$, 
while the SM quarks are triplets under the first one.
The coloron decay into a pair of vectorlike quarks has a width \cite{Dobrescu:2009vz}
\be
\Gamma (G' \to \chi \bar \chi)  =   \dfrac{1}{6 \tan^4\! \theta} \;   \left( 1 + \dfrac{2 m_\chi^2 }{M_{G'}^2 } \right)\left( 1 - \dfrac{4 m_\chi^2 }{M_{G'}^2 } \right)^{\! 1/2} \,  \sum_q \Gamma (G' \to q\bar q)  ~~.
 \label{eq:chi2}
\ee
As this is increasing the total  width of the coloron, the range of $\tan\theta$ consistent with
$\Gamma_{G'} / M_{G'} < 7\% $ becomes $0.5 \lesssim \tan\theta  \lesssim 0.8$. The width-to-mass ratio of the coloron is plotted in Figure \ref{fig:GpWidth}.

\begin{figure}[t]
\hspace*{-1mm} \includegraphics[width=0.47\textwidth]{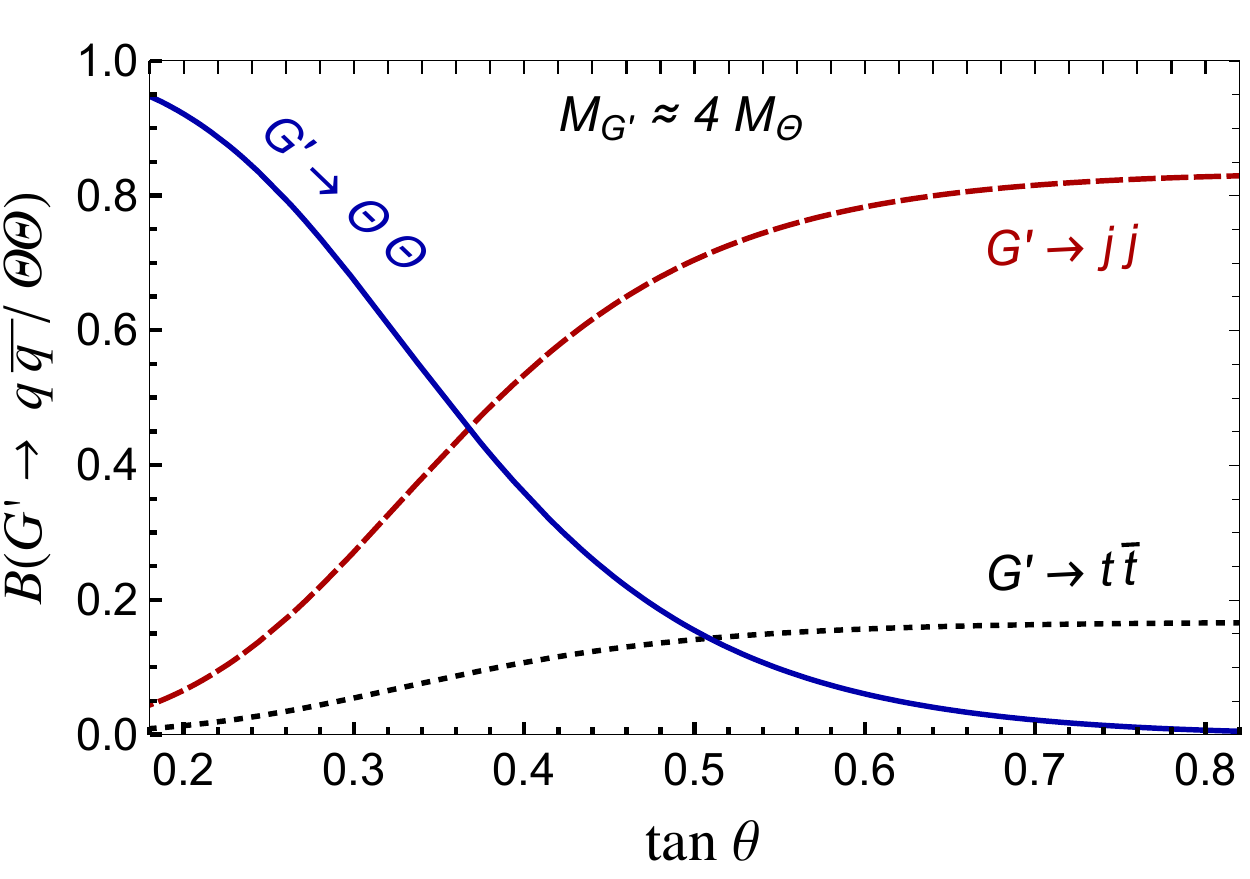} \hspace*{4mm}  \includegraphics[width=0.47\textwidth]{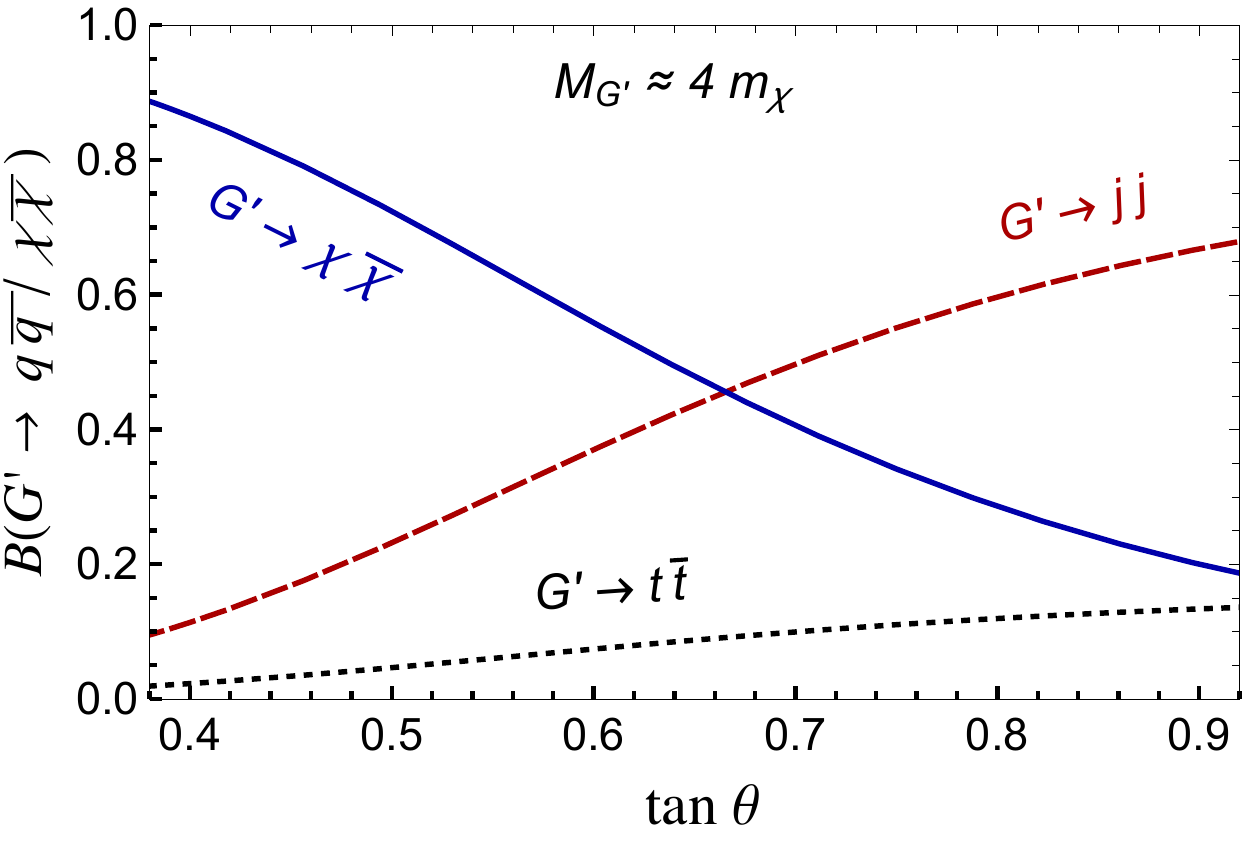} 
\vspace*{-1mm}
\caption{Branching fractions of the coloron as a function of the coupling normalization $\tan\theta$. 
The dijet branching fraction  (dashed red line) is summed over five $q\bar{q}$ flavors.
{\it Left panel:} Coloron+scalar model with $M_{G'} /M_\Theta \approx 4$.
The branching fraction into color-octet scalars (solid blue line) vanishes at $\tan\theta =1$. 
{\it Right panel:} Coloron+quark model with $M_{G'} /m_\chi  \approx 4$. 
The sensitivity to the mass ratio is low, except near threshold ($M_{G'} =2 M_\Theta$ or $M_{G'} =2 m_\chi$). 
 \\ [-3mm] }
\label{fig:BR}
\end{figure}

These partial widths for $G' \to \Theta \Theta$ and  $G' \to \chi \bar \chi$, together with the $q\bar q$ width given in Eq.~(\ref{eq:qqbarWidth}) lead to the coloron 
branching fractions shown in 
Figure \ref{fig:BR}. The coloron mass is much larger than the top quark mass, so the branching fraction into quark jets is five times larger than that into a top quark pair.  The masses of the color-octet scalar or of the vectorlike quark are closer to $M_{G'}$,
so that the phase-space suppression factors in Eqs.~(\ref{eq:Theta2}) and  (\ref{eq:chi2}) cannot be neglected. 
In Figure \ref{fig:BR} the mass ratios $M_\Theta/M_{G'}$ and
$m_\chi/ M_{G'}$ are fixed at 1/4, but for smaller ratios the plots would not change significantly. For larger mass ratios, the decay of the coloron into new particles has smaller branching fractions.

In the coloron+scalar model,
the $\Theta$ scalar decays into two gluons with a branching fraction of nearly 100\%. This decay occurs at one loop, with contributions 
from both a coloron loop  and a $\Theta$ loop  \cite{Bai:2018jsr}, so the width of $\Theta$ is very narrow (nevertheless, the decay is prompt).
Thus, the $s$-channel production of a coloron followed by the cascade decay $G' \to \Theta\Theta \to (gg)(gg)$ leads to a pair of dijets (see left diagram of Figure \ref{fig:diagrams}).
Each dijet has a mass given by $M_\Theta$, and the invariant mass distribution of the four jets has  a peak at $M_{G'}$.

For $M_{G'} \gg M_\Theta$ each dijet would appear as a single wide jet with a 2-prong substructure. Consequently, the coloron signal can show up in a dijet resonance search, with the special feature that
requiring each wide jet to have substructure can eliminate most of the background.

Given that the $\Theta$ scalar carries color, its couplings to gluons lead to nonresonant production of two $\Theta$'s. 
The process $p p \to \Theta\Theta \to (gg)(gg) $ has a larger cross section than $p p \to G' \to \Theta\Theta \to (gg)(gg) $ 
when $M_\Theta \ll M_{G'}$, but as discussed in Section \ref{sec:signalDiquark}
its contribution to the signal of an ultraheavy coloron is negligible.  
The nonresonant $\Theta\Theta$ production leads though to a lower limit on the scalar mass. 
The cross section for the nonresonant $p p \to \Theta\Theta \to (jj)(jj) $ process at the 13 TeV LHC, including the QCD NLO corrections which induce an increase of about 80\% 
 \cite{GoncalvesNetto:2012nt}, 
 is approximately 130 fb for $M_\Theta = 0.9$ TeV, and 50 fb for $M_\Theta = 1$ TeV.    
The recent CMS search  \cite{Sirunyan:2018rlj} for a pair of dijet resonances sets a limit $M_\Theta \gtrsim 0.95$ TeV.

\begin{figure}[t]
\hspace*{-1mm} \includegraphics[width=0.46\textwidth]{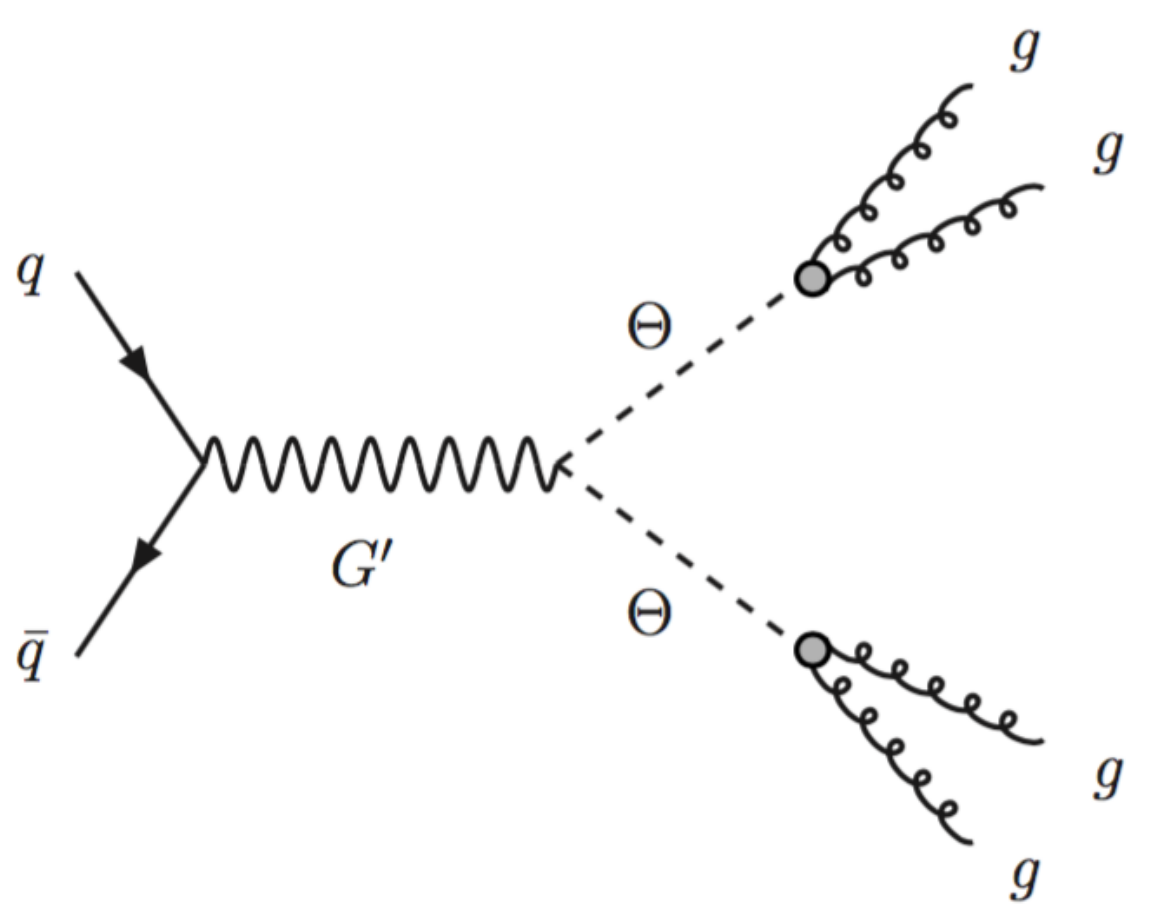}   \hspace*{10mm}   \includegraphics[width=0.46\textwidth]{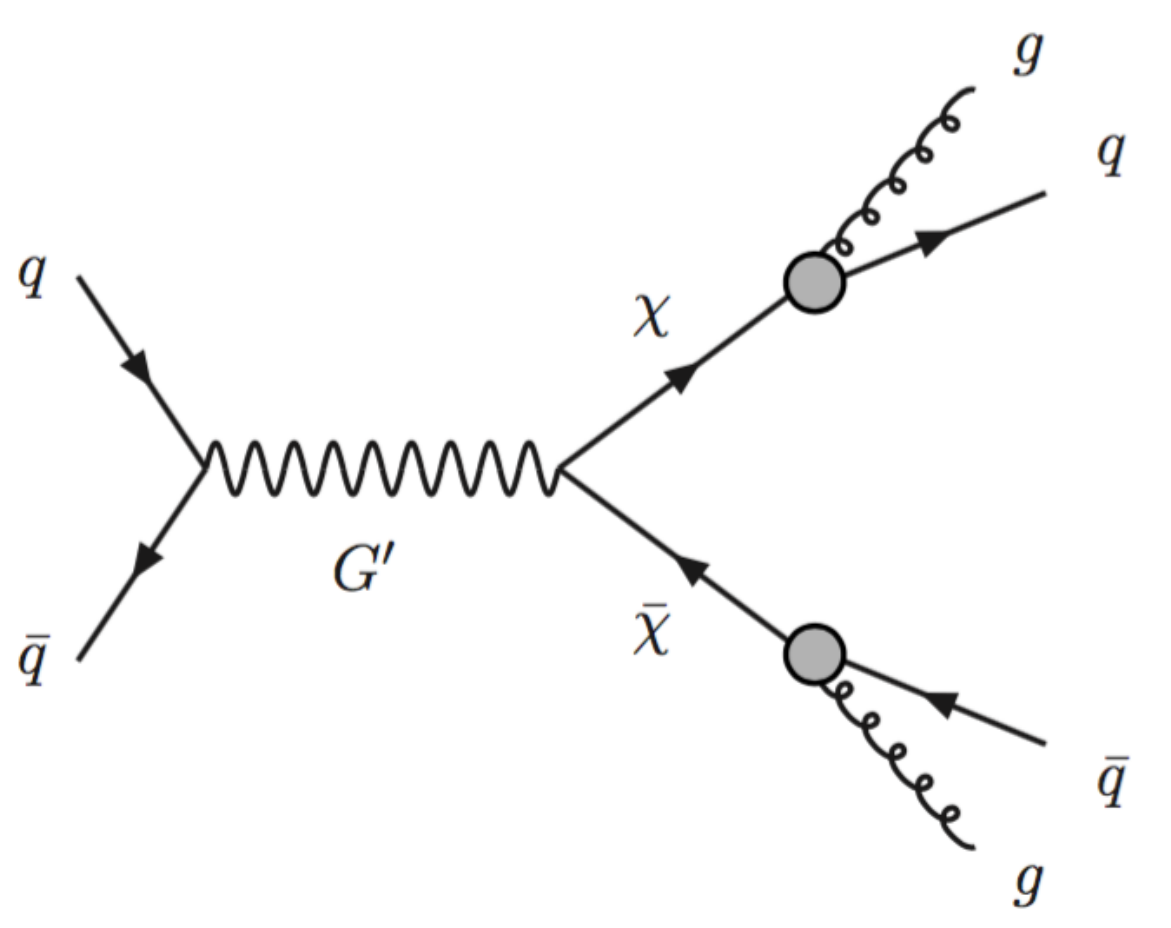}
\vspace*{-9mm}
\caption{Coloron production followed by decay into a pair of color-octet scalars (left diagram) or vectorlike quarks  (right diagram), each of them then decaying at one loop into two hadronic jets.
 \\ [-3mm] }
\label{fig:diagrams}
\end{figure}

In the coloron+quark model, there are various possibilities for the decay of the vectorlike quark $\chi$, as discussed in Section 
\ref{sec:signalDiquark}.
For a simpler comparison with the coloron+scalar model, we assume here that $\chi$ decays into two jets with 
a branching fraction of nearly 100\%. This decay may arise from the $\chi \to u g$ (or $c g$) process, where the up (or charm) quark and the 
gluon are attached to a loop involving some very heavy colored particles. 
An alternative is the $\chi \to u \,\phi_R$ (or $c \,\phi_R$)   process at tree level, with the  $\phi_R$ scalar decaying into 
two gluons; if $\phi_R$ is much lighter than $\chi$, then the two gluons are collimated and $\phi_R$ appears as a single jet.
In either case, the $G' \to \chi\bar \chi \to (jj)(jj)$ process leads to resonant production of a pair of  equal-mass dijet resonances.
For $M_{G'} \gg m_\chi$, the signal is again a pair of wide jets, each with 2-prong substructure (see right diagram of Figure \ref{fig:diagrams}).

The QCD background computed in Section \ref{sec:backDiquark} (see Figure~\ref{fig:QCDxsec}), 
summed over a number of bins  (each of 100 GeV) given by $0.8 M_{G'}/1$ TeV
is about an order of magnitude 
smaller than the coloron production cross section shown in Figure~\ref{fig:xsec}. 
Multiplying the latter by $\tan^2\theta$ and by the branching fraction for $G' \to \Theta\Theta$ or $G' \to \chi\bar \chi$ shown in Figure \ref{fig:BR}, and by the signal efficiency with the cuts listed in Section \ref{sec:backDiquark}, 
we find that the number of signal events is larger than the number of background events as long as  $\tan\theta$ is above  0.4 (and not near 1 in the case of the coloron+scalar model).
For $\tan\theta \approx 0.6$, a coloron as heavy as  8.5 TeV can be discovered through the  $G' \to \chi\bar \chi \to (jj)(jj)$ process by the end of the high-luminosity run of the LHC.

\bigskip

%\newpage
%%%%%%%%%%%%%%%%%%%%%%%%%%%%%%%%%%%%%%%%%%%%%%%%%%%%%%%%%%%%%%%
%%%%%%%%%%%%%%%%%%%%%%%%%%%%%%%%%%%%%%%%%%%%%%%%%%%%%%%%%%%%%%%%%%%%%%%%%%%%
\section{Events with a mass of 8 TeV}
\label{sec:8TeV}
\setcounter{equation}{0}

Experiments at the LHC have recently conducted searches for new particles in both the dijet~\cite{CMS-PAS-EXO-17-026,Sirunyan:2018xlo,Aaboud:2017yvp} 
and 4-jet final states~\cite{Sirunyan:2018rlj,Aaboud:2017nmi}. In this section we explore some hints of an ultraheavy resonance in the data from 
the dijet resonance searches, and discuss the sensitivity of the 4-jet searches.

%%%%%%%%%%%%%%%%%%%%%%%%%%%%%%%%%%%%%%%%%%%%%%%%%%%%%%%%%%%%%%%%%%%%%%%%%%%%%%%%%%%%%
\subsection{Dijet and 4-jet LHC searches and events}
\label{sec:event8TeV}

The CMS experiment has found an event which is a candidate for an ultraheavy resonance decaying into a pair of 
massive particles. It was reported in the most recent CMS dijet resonance search using 77.8 fb$^{-1}$ of integrated 
luminosity from 13 TeV $pp$ collisions~\cite{CMS-PAS-EXO-17-026}.  The properties of the event, as given in \cite{CMS-PAS-EXO-17-026}, 
are listed again in Table~\ref{tab:event}. There are four distinct jets in the event reconstructed with 
the standard jet algorithm used by CMS, the anti-$k_T$ clustering algorithm~\cite{Cacciari:2008gp} with a distance parameter $R=0.4$. 
The dijet resonance search at CMS uses wide jets (labelled here by $J$) to reconstruct the dijet mass of the event. The two wide jets are seeded by the 
corresponding two highest $p_T$ standard jets, and additional jets are included in the two wide jets if they
are within a distance $\Delta R_J =\sqrt{\smash[b]{(\Delta\eta_J)^2 +  (\Delta\phi_J)^2}}<1.1$.  The wide jet algorithm is used by CMS to improve
the dijet mass resolution in the presence of final state radiation, but it also can reconstruct an ultraheavy resonance decaying 
to pairs of masive secondary particles.  The ultraheavy resonance decay gives a pair of wide jets, where each wide jet is composed of two standard jets
resulting from the decay of the secondary particle.  The event in Table~\ref{tab:event} has a dijet mass $m_{JJ} \approx 8.0$ TeV, the highest value from 
CMS so far. In addition to the high dijet mass, the event has a very low 
probability of being produced by the QCD background, because it has four standard jets, and furthermore each pair of standard jets reconstructs 
to a large mass of 1.8 TeV. In Section \ref{sec:QCD8TeV} we calculate that the QCD background is below $10^{-4}$ events.

\begin{table}[t]
\begin{center}
\begin{tabular}{|ccc|ccc|cccc|}
\hline
\multicolumn{3}{|c|}{Dijet Event} & \multicolumn{3}{c|}{Wide Jets} & \multicolumn{4}{c|}{Standard Jets} 
\\ \hline 
 \hspace*{0.1cm} Mass \hspace*{0.1cm} &  \hspace*{0.1cm} \multirow{2}{*}{$\Delta\eta_{JJ}$}   \hspace*{0.1cm} &  \multirow{2}{*}{$\Delta\phi_{JJ}$}   \hspace*{0.1cm}  &  \hspace*{0.1cm} $p_{TJ}$ \hspace*{0.1cm} & \hspace*{0.1cm} Mass \hspace*{0.1cm} &  \hspace*{0.1cm}  \multirow{2}{*}{$\Delta R_J$} \hspace*{0.1cm} & \hspace*{0.2cm}  \multirow{2}{*}{index}  \hspace*{0.2cm}  & $p_{Tj}$ & \multirow{2}{*}{$\eta_j$} & \multirow{2}{*}{$\phi_j$} 
  \\ [-2mm]
(TeV) &               &                 & \hspace*{0.2cm} (TeV) \hspace*{0.2cm} & (TeV)&   &    & (TeV) &     &           \\ \hline  \hline
\multirow{4}{*}{8.0} &  \multirow{4}{*}{0.4} & \multirow{4}{*}{3.1} & \multirow{2}{*}{3.5} &  \multirow{2}{*}{1.8}  &   \multirow{2}{*}{0.98}  & 0 & 2.16 & 0.27 & 1.47   \\ 
 &  &  &  & &         & 2 & 1.68 & 0.21 & 2.45 \\ \cline{7-10}
 & & &  \multirow{2}{*}{3.4} &  \multirow{2}{*}{1.8} &  \multirow{2}{*}{1.03} & 1 & 1.99 & 0.29 & $-1.27$   \\ 
 &  &  &  &   &       & 3 & \hspace*{0.2cm} 1.40 \hspace*{0.2cm} & \hspace*{0.2cm} $-0.74$ \hspace*{0.2cm} &  \hspace*{0.1cm} $-1.17$ \hspace*{0.1cm} \\
\hline
\end{tabular}
\end{center}
\caption{Properties of the CMS \cite{CMS-PAS-EXO-17-026}  
highest-mass dijet event, which contains two wide jets, each made from two standard jets. 
The quantities of the dijet event are the invariant mass, pseudorapidity separation ($\Delta\eta_{JJ}$), and azimuthal separation ($\Delta\phi_{JJ}$) of
the pair of wide jets. The quantities of the wide jets are the transverse momentum ($p_{TJ}$), the invariant mass ($m_J$), and the separation ($\Delta R_J$)  of the pair of standard jets
within each wide jet.
The quantities of the standard jets are the $p_{Tj}$-ordered index,  the transverse momentum ($p_{Tj})$, pseudorapidity ($\eta_j$), and azimuthal angle ($\phi_j$). \\ [-3mm] }
\label{tab:event}
\end{table}

 The CMS experiment has searched for signals of massive dijet resonances and has reported events which are candidates for a new particle of mass near 8 TeV. 
 In addition to the 4-jet event at 8.0 TeV, the CMS search \cite{CMS-PAS-EXO-17-026}   using 77.8 fb$^{-1}$ also reported a typical dijet 
event with a mass of 7.9 TeV in a dijet mass bin of width $0.3$ TeV, giving a total of 2 events observed in that bin. The estimated QCD background from PYTHIA 8.2 \cite{Sjostrand:2014zea}
in that bin is 0.6 events. That bin width corresponds to the CMS resolution for a dijet resonance at that 
mass, so the majority of a signal would generally appear in roughly 2 bins, corresponding to the mass region $7.7<m_{jj}<8.3$ TeV.  In that region
CMS observes 2 events while the background is about 1.1 events, which is a conservative estimate because the background is mainly coming from the 
lower edge of the mass region not directly underneath the peak.
The actual local significance of a dijet resonance signal at 8.0 TeV in that dataset reported by CMS
is 1.2 standard deviations. This small but positive signficance arises because of the peaking of the
signal at 8 TeV, the falling shape of the background, and the inclusion of both the dijet event at 7.9 TeV and the event with the 4-jet topology at 8.0 TeV.   

The ATLAS experiment has also searched for massive dijet resonances and has also observed events which are candidates for a new particle 
of mass near 8 TeV decaying to $jj$~\cite{Aaboud:2017yvp}.  That search uses only 37 fb$^{-1}$ of integrated luminosity, and the two highest-mass events from ATLAS have a dijet mass 
of approximately 8.0 and 8.1 TeV. The data from ATLAS is more suggestive of a signal than the data from CMS because those two events are on the 
tail of the distribution separated from the lower mass events by a gap of about 1 TeV.  The ATLAS events 
are in two adjacent bins, each of width about $0.1$ TeV, with a QCD background estimate of only 0.2 events total for those two bins.
While 2 events on a background of 0.2 events
appears significant, we note that the majority of a resonance signal at ATLAS would appear in more bins than 
at CMS.\footnote{This is because the ATLAS binning corresponds to their calorimeter resolution for the particles inside the dijet, while the CMS binning 
corresponds to the wider resolution for a dijet resonance, for which the two final state partons can radiate before they hadronize.} \
Conservatively assuming a similar dijet mass resolution from a resonance for ATLAS as for CMS, we estimate from \cite{Aaboud:2017yvp}  that the
background within the dijet mass interval $7.7<m_{jj}<8.3$ TeV for ATLAS is 0.5 events, again coming mainly from the lower edge of the
mass interval.  So the indications for a dijet resonance at ATLAS at 8 TeV are stronger than at CMS. It would be useful to 
have an estimate of the local significance of a narrow dijet resonance at 8 TeV from this ATLAS dataset.
We note that the ATLAS search for a dijet resonance would not contain a 4-jet event from an 8 TeV resonance, 
like in Table~\ref{tab:event}, because ATLAS does not use wide jets and instead uses standard jets from the anti-$k_T$ clustering  algorithm with 
a distance parameter $R=0.5$. 

The CMS and ATLAS experiments have conducted searches in the 4-jet final state for signals with pairs of dijet resonances.
The searches are not very sensitive to a signal from an 8 TeV resonance decaying to secondary particles which in turn decay to dijets.
The CMS experiment has conducted a search for pair-produced resonances decaying to quark pairs using 35.9 fb$^{-1}$ of integrated 
luminosity~\cite{Sirunyan:2018rlj}.   The sample 
of events was recorded during the 2016 run and does not contain the event of Table~\ref{tab:event} which came from the 2017 run.
The distribution of the average dijet mass of the two pairs, 
$\overline{m}_{jj} = (m_{jj1}+m_{jj2})/2$, falls steeply and the highest observed value is $\overline{m}_{jj}\approx 1.6$ TeV, 
smaller than the value $\overline{m}_{jj}=1.8$ TeV for the event in Table~\ref{tab:event}. 

A search for pair-produced dijet resonances  from the ATLAS experiment using 36.7 fb$^{-1}$ of integrated luminosity~\cite{Aaboud:2017nmi} presents a 
distribution of $\overline{m}_{jj}$, and a table of signal and background. This search observes 2 events with an expected background 
of about 2 events for average dijet mass values $\overline{m}_{jj}>1.79$ TeV. Neither the CMS nor the ATLAS searches indicate the 
observed 4-jet mass, so we cannot tell if any of the events have a 4-jet mass of 8 TeV. The CMS and ATLAS searches integrate over 
the 4-jet mass and search for signals using only $\overline{m}_{jj}$, resulting in a much larger background from QCD than would be expected
near a 4-jet mass of 8 TeV. Therefore, the reported 4-jet search data from the LHC cannot be used to estimate the QCD background to the event 
in Table~\ref{tab:event}. 

The CMS and ATLAS 4-jet searches were optimized to search for nonresonant pair production of some new colored particle
from its QCD couplings, such as a coloron pair \cite{Dobrescu:2007yp,Buschmann:2017ucg}, or a pair of  colored scalars  \cite{Chivukula:1991zk,Dobrescu:2007yp}.
We propose here a more general search for pair production of dijet resonances at the LHC, 
exploring the distribution of the 4-jet mass as a function of the average dijet mass of the pairs,
which would be sensitive to s-channel production of an ultraheavy resonance.

%%%%%%%%%%%%%%%%%%%%%%%%%%%%%%%%%%%%%%%%%%%%%%%%%%%%%%%%%%%%%%%%%%%%%%%%
\subsection{QCD background to resonant production of a pair of dijets}
\label{sec:QCD8TeV}

We next estimate the QCD background for the event in Table~\ref{tab:event}.
The most restrictive properties of the event are its large 4-jet invariant 
mass, $m_{4j} =8$ TeV, and the large mass of each of the two wide jets in the 
event, $m_J = 1.8$ TeV. Therefore we calculate the probability of getting an event 
from QCD with masses equal to or exceeding these values, along with the other 
cuts imposed by the CMS search. 

The QCD 4-jet production at large masses was  generically  discussed 
in Section \ref{sec:backDiquark}. The main contributions arise from quark-quark initial states which radiate two gluons, as shown in Figure~\ref{fig:QCDdiagrams}.
We generate the QCD $pp \to 4j$ process using MadGraph5\_aMC@NLO~\cite{Alwall:2014hca}, 
run at LO with $\sqrt{s} = 13$ TeV, using the NNPDF2.3LO~\cite{Ball:2017nwa} parton distributions. The 4-jet cross section has been calculated to NLO in \cite{Bern:2011ep},
and the $K$-factor was found to be  close to one for  large transverse momenta. Therefore, we do not expect the NLO corrections to significantly change the
predicted QCD background. Additionally, we expect the effects of the parton shower or hadronization not to exceed  10\%, since the observables discussed below 
are not sensitive to soft or collinear radiation. We will neglect these effects in what follows.

Following the CMS search selection, we apply the CMS wide-jet algorithm to the four 
final state partons in the event, which are required to be separated by 
$\Delta R>0.4$ to mimic the standard anti-$k_T$ algorithm.  We also
require the two wide jets to be separated in pseudorapidity by 
$|\Delta\eta_{JJ}|<1.1$. We produced two
samples of $3\times 10^4$ events: one with tight cuts and one with loose cuts.
For both samples we additionally place requirements on jet transverse momentum and pseduorapidity 
which are fully efficient for the QCD background at the large masses discussed here. 

The QCD background for the CMS event is computed from the $pp \to 4j$ sample with tight
cuts: $m_{4j} \ge 8$ TeV, $m_{J1} \ge 1.8$ TeV and $m_{J2} \ge 1.8$ TeV. The QCD
cross section is $5.7\times 10^{-7}$ fb, resulting in $4.5\times 10^{-5}$ events
expected in 77.8 fb$^{-1}$ of data. 
Performing PDF reweighting we find that for the CT14  set \cite{Dulat:2015mca} at LO the 
cross section is $6.2\times 10^{-7}$ fb corresponding to $4.9\times 10^{-5}$ events, while for the
MMHT2014  set \cite{Harland-Lang:2014zoa} at LO
the cross section is $4.8\times 10^{-7}$ fb resulting in  $3.8\times 10^{-5}$ events.
We conclude that the Poisson probability of observing the event in Table~\ref{tab:event} from the QCD 
background alone is approximately $5 \times 10^{-5}$, equivalent to a 4 standard 
deviation excursion of the background.

The QCD cross section has a PDF uncertainty of 21\% from NNPDF2.3LO.  Note that this uncertainty is relatively small 
because the dominant contribution only  involves valence quarks. The MadGraph simulation also reports that this LO
cross section varies by $^{+72\%}_{-40\%}$ when the renormalization $(\mu_R)$ and factorization $(\mu_F)$ scales are
varied by a factor of 2, according to the seven scale scheme, {\it i.e.}, by finding the maximum excursion in the cross section among the following seven choices of the two scales:
$(\mu_F/ m_{4j} ,\mu_R/ m_{4j})=(0.5,0.5),(1,0.5),(0.5,1),(1,1),(1,2),(2,1),(2,2)$.

%%%
\begin{figure}[t]
%\vspace*{3mm}
\hspace*{-1mm}\includegraphics[width=0.489\textwidth]{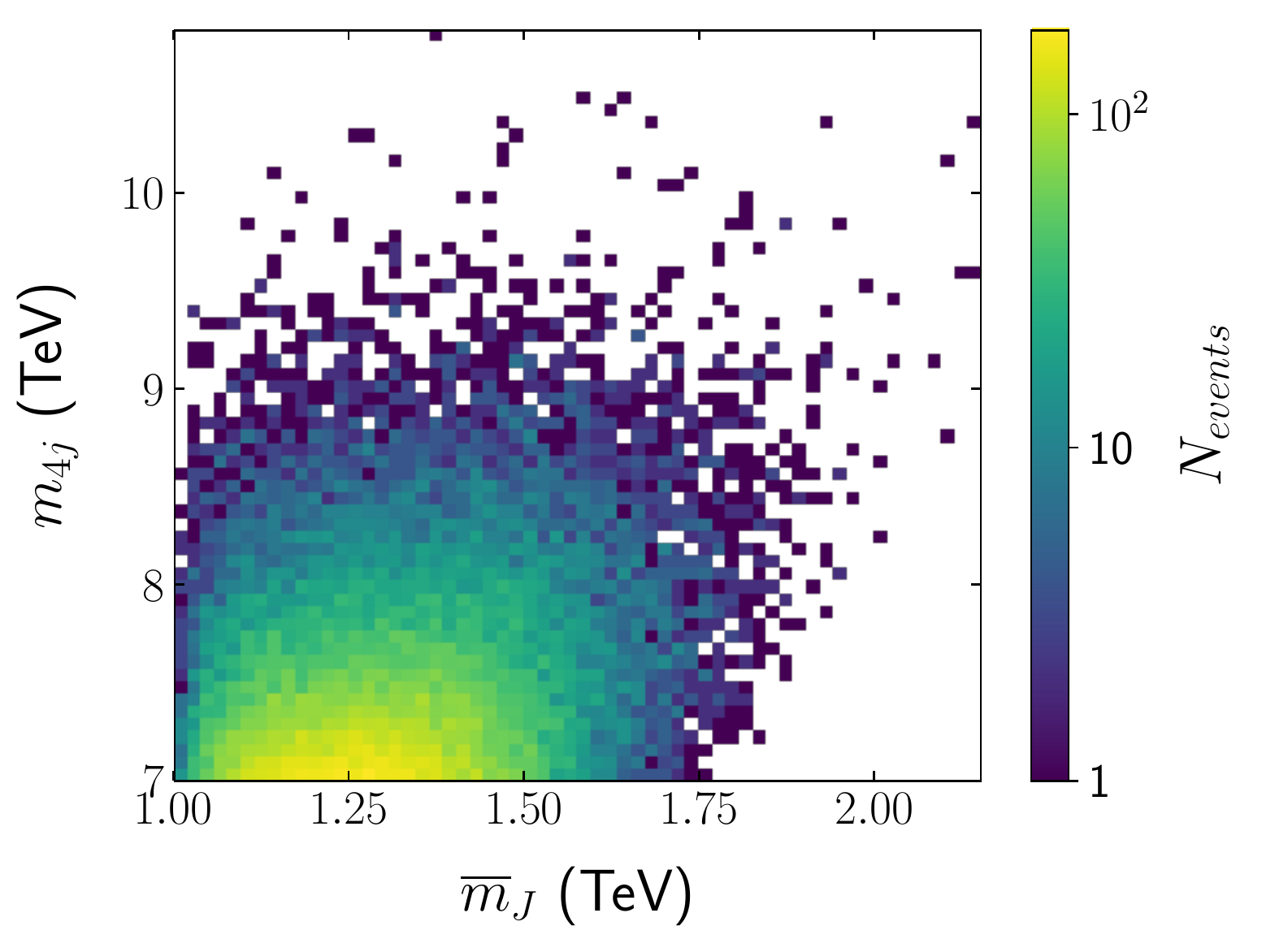}   \hspace*{1mm}
\includegraphics[width=0.497\textwidth]{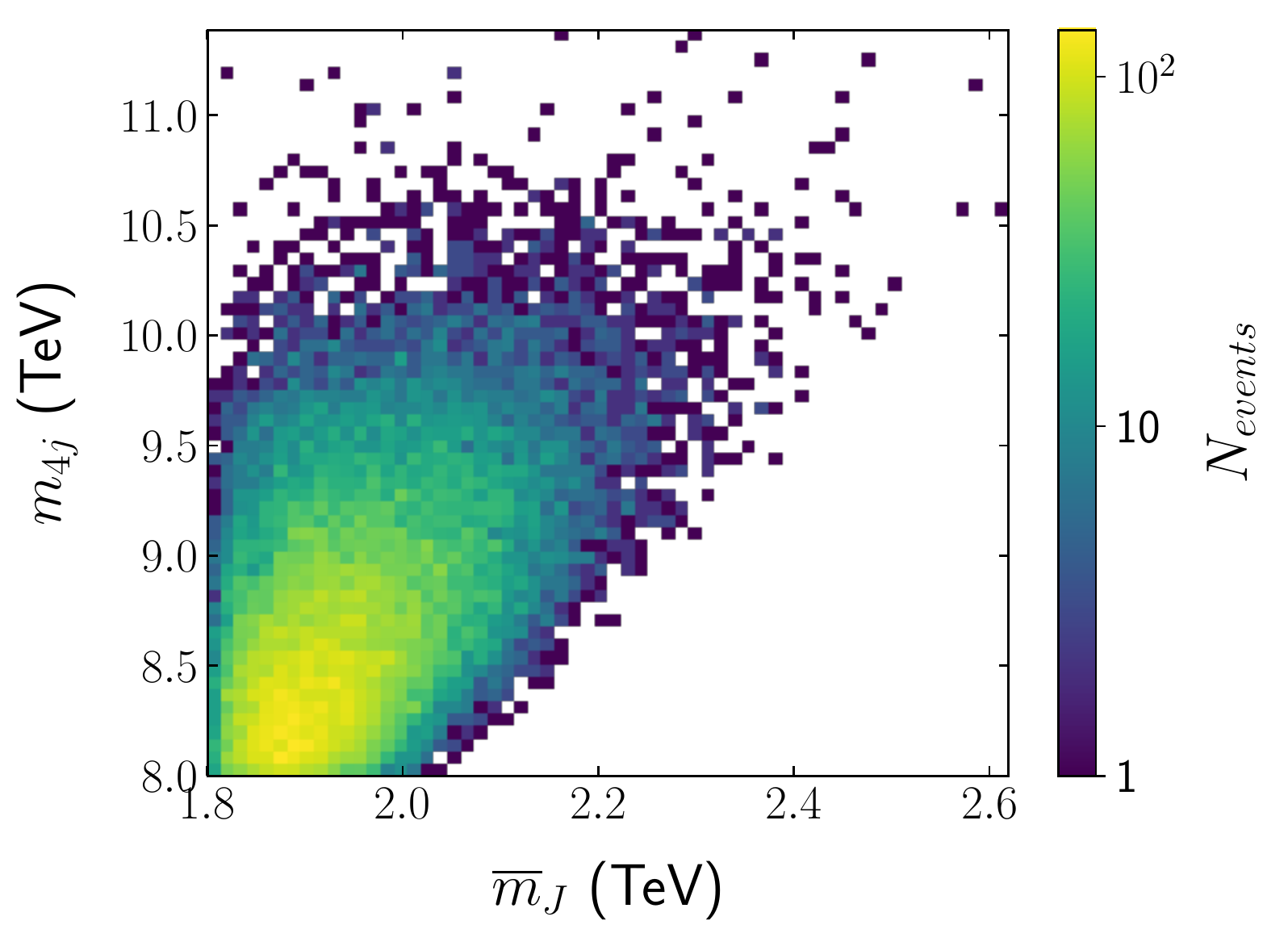}  

 \vspace*{-5mm}
\caption{ QCD distribution of the four parton invariant mass ($m_{4j}$) versus the average diparton mass for the two pairs 
of partons ($\bar m_J$), from the MadGraph LO samples with the loose cuts (left panel) or  tight cuts (right panel) described in the text. 
\\ [-1mm]  }
\label{fig:M4jvMavg}
\end{figure}
%%%

The QCD distributions are studied in the tight sample and also a sample 
with loose cuts: $m_{4j}>7$ TeV, $m_{J1}>1$ TeV and $m_{J2}>1$ TeV.  In Figure~\ref{fig:M4jvMavg} 
we show the distribution of $m_{4j}$ vs. $\overline{m}_J=(m_{J1}+m_{J2})/2$. 
The distribution falls off rapidly
with increasing  $m_{4j}$ and $\overline{m}_J$. We see 
that background events with high values of both $m_{4j}$ and $\overline{m}_J$ are more rare than events with
high values of only one of the two masses, so a search for a massive particle decaying to pairs of 
dijet resonances using both $m_{4j}$ and $\overline{m}_J$ would be more sensitive than the existing searches
at the LHC which have used only one of the two masses.

The diagonal edge to the distributions in Figure~\ref{fig:M4jvMavg}, {\it i.e.,} the 
maximum allowed value of $\overline{m}_J$ which increases with $m_{4j}$, 
originates from the kinematics of the $\Delta R_J <1.1$ requirement for the two partons to be combined into a wide jet.
That requirement forces the wide jet mass $m_J$ to be proportional 
to the wide jet $p_T$, because $m_J \approx p_{TJ}\, \Delta R_J/2$ for central jets.
However the $p_T$ of the wide jets is limited at a fixed $m_{4j}$, creating a maximum 
value of $\overline{m}_J$ for any given value of $m_{4j}$. In the case where the two wide jets have equal mass, 
the following approximate relation holds for central jets
\begin{equation}
\overline{m}_J\approx \frac{m_{4j}}{2\sqrt{1+4/\Delta R_J^2}}  ~~~~.
\label{eq:kinematics}
\end{equation}
For $m_{4J}=8$ TeV and a maximum value $\Delta R_{J \rm max}=1.1$, we find the maximum allowed value of wide jet mass is
$m_{J_{\rm max}}\approx 1.9$ TeV, from the kinematics estimate alone, which is consistent with the maximum value shown
in Figure~\ref{fig:M4jvMavg},  $m_{J_{\rm max}} \approx 2$ TeV. Therefore, the wide jet mass of 1.8 TeV for the CMS event 
in Table~\ref{tab:event} is close to the upper boundary of what is kinematically allowed for the observed 
4-jet mass of 8.0 TeV given the wide jet contraint $\Delta R_J<1.1$.  This is similar to the information that 
the separation of the standard jets in each pair in Table~\ref{tab:event} is $\Delta R_{02}=0.98$ and 
$\Delta R_{13}={1.03}$, close to the maximum allowable ($\Delta R_{J \rm max}$) for the wide jet algorithm.

%%%%%%
\begin{figure}[t]
\hspace*{-1mm}\includegraphics[width=0.487\textwidth]{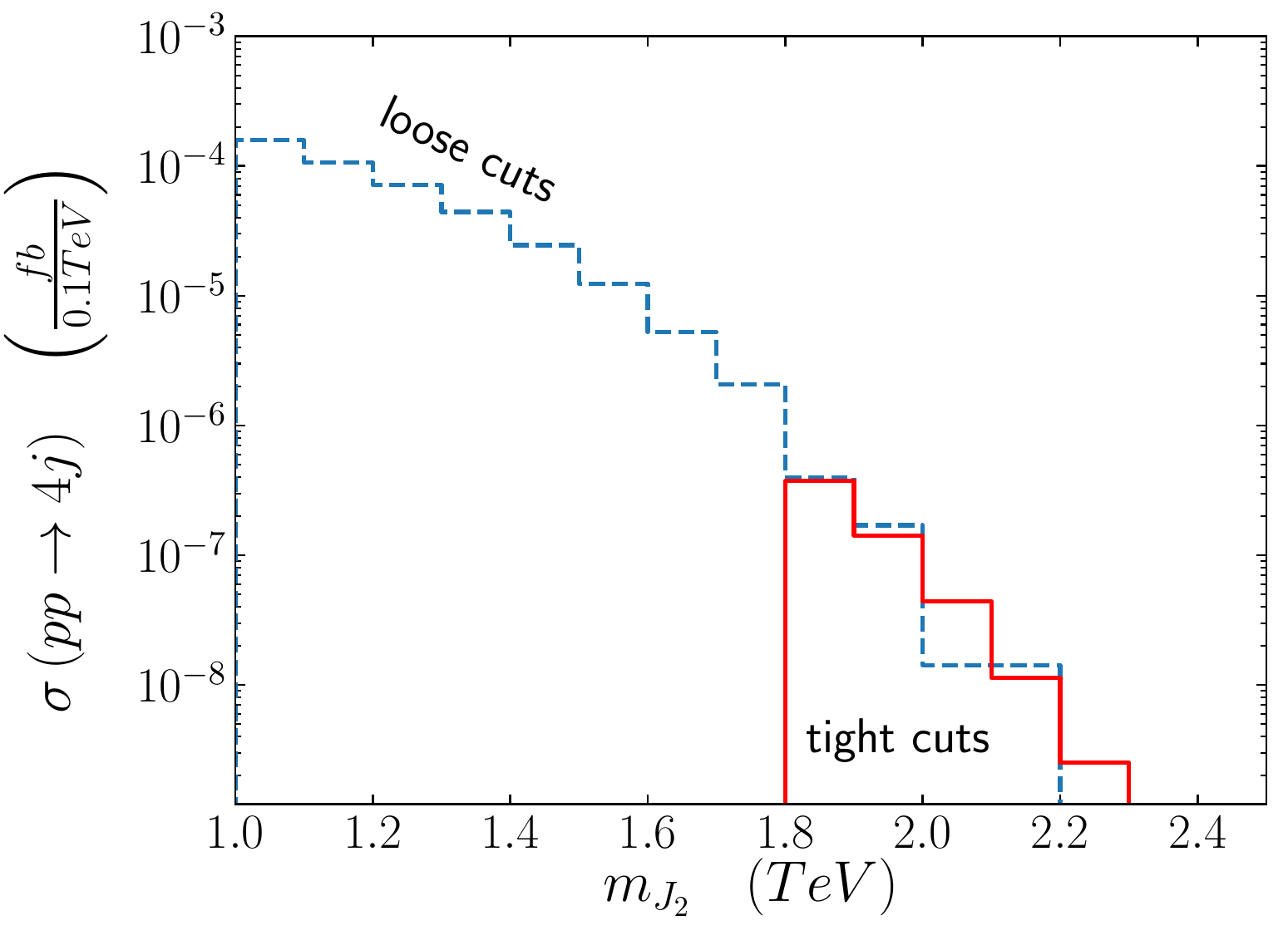}    %.eps} 
\hspace*{1mm}
\includegraphics[width=0.499\textwidth]{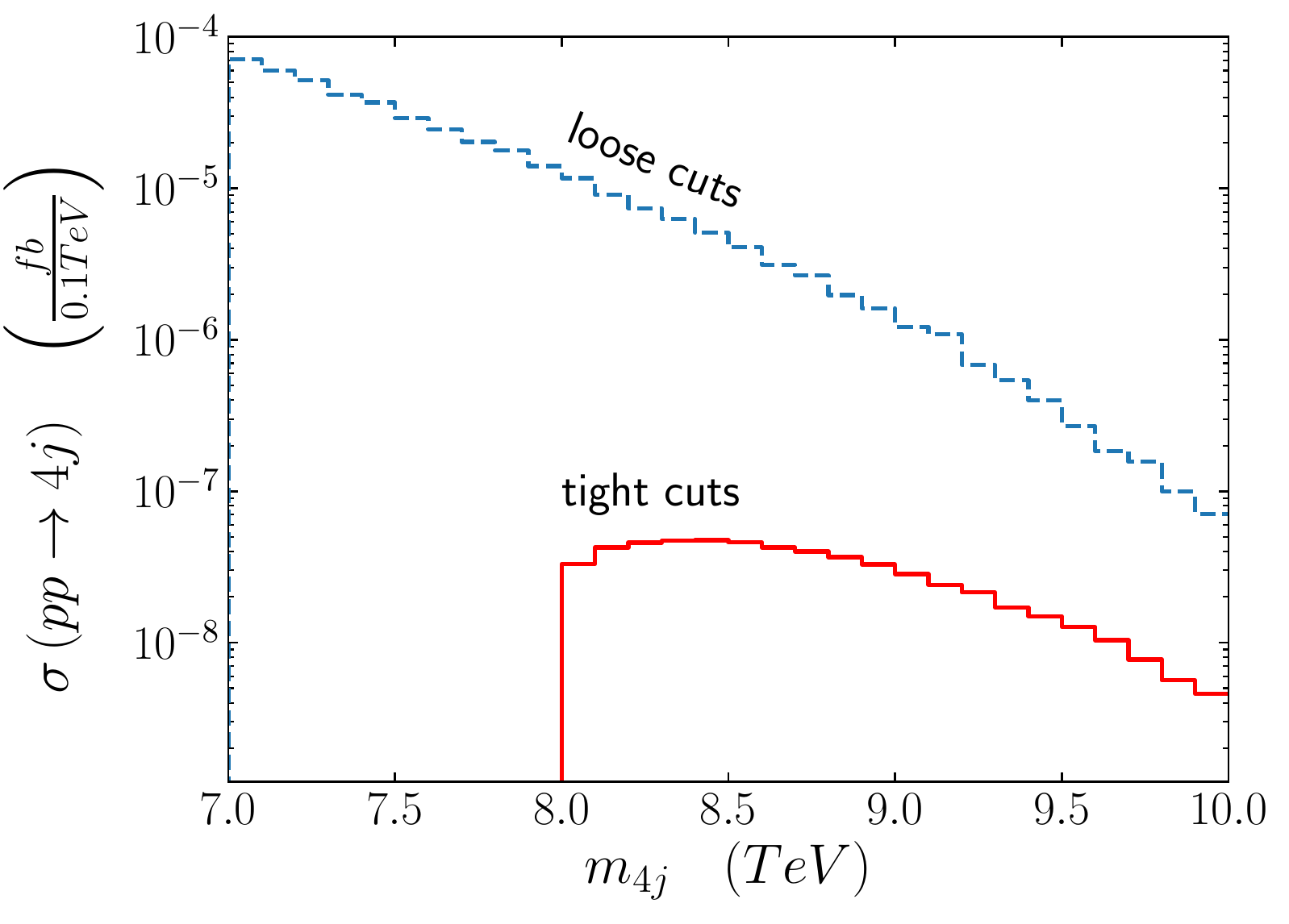}    %eps}    

\vspace*{-4mm}
\caption{QCD cross section (in fb/bin)  for the  $pp \to 4j$ process  at $\sqrt{s} = 13$ TeV 
computed with MadGraph at LO,  as a function of  the smaller wide jet mass $m_{J2}$ in the event (left panel),  
or of  the 4-parton mass $m_{4j}$ (right panel).
The results are obtained with the loose cuts (blue dashed line) or  tight cuts (red solid line) described in the text. \\ [-2mm] }
\label{fig:MJ2andM4j}
\end{figure}
%%%%%%

In Figure~\ref{fig:MJ2andM4j} we show distributions of the wide jet mass, $m_{J2}$, which is the 
smaller of the two masses, and distributions of $m_{4j}$.
The cross section as a function of $m_{J2}$ falls significantly faster than $m_{4j}$.  In the
tight sample, the  $m_{J2}$ distributions fall off with little bias from the $m_{4j}>8$ TeV requirement,
while the $m_{4j}$ distribution is flattened at low mass by the $m_{J1}, m_{J2}>1.8$ TeV requirements.  
We see that the latter are more restrictive than the $m_{4j}>8$ TeV 
requirement, because QCD events with $m_{J1}, m_{J2}>1.8$ TeV are more rare than those with  $m_{4j}>8$ TeV when the requirement $\Delta R_J < 1.1$ is imposed.  In the loose
sample, which has the requirements $m_{J1}, m_{J2}>1$ TeV, there is no visible bias in the $m_{4j}$ distribution which
falls rapidly as a function  of $m_{4j}$, confirming that the $m_{J1}, m_{J2}>1.8$ TeV cut caused 
the bias in  $m_{4j}$ in the tight sample. This is further supported by the fact that the cross sections for the tight and loose samples for $m_{J2}>1.8$ TeV are about the same, again demonstrating that the 4-jet invariant mass cut has a small effect as compared to the wide jet mass cut.

The distributions of the wide jet mass $m_{J1}$, which is the 
larger of the two masses, and of $m_{J2}$ vs. $m_{J1}$ are displayed in  Figure~\ref{fig:MJ1andMJ2}.
The wide jet mass $m_{J1}$ also falls rapidly, like $m_{J2}$, but shows a small bias near the $m_{J1}=1.8$ TeV cut value 
caused by requiring $m_{J1}>m_{J2}$.  We can see that the larger value of wide jet mass $m_{J1}$ is expected to be within a few hundred GeV of
the smaller value of wide jet mass $m_{J2}$.  We note that the dijet resonance mass resolution at CMS is 
about 100 GeV for a $1.8$ TeV dijet mass. Therefore, the two wide jets are expected to have similar mass values   
within the expectations from experimental resolution, even when they originate from the QCD background. So 
the coincidence of observing both wide jets to have a mass of 1.8 TeV in Table~\ref{tab:event} is not as 
unusual as the large value of that mass. Further, as long as the pairs of jets are constrained to be within the wide jet cone, we do not expect that much enhancement of signal sensitivity 
could be achieved in an analysis by requiring the two pairs of jets to have the same mass.  
Loosening the wide jet requirement $\Delta R_J <1.1$ on the pairs of jets greatly increases the signal
efficiency, while the QCD background would still be negligible for the large masses discussed here.

%%%
\begin{figure}[t]
%\vspace*{2mm}
\hspace*{-1mm}\includegraphics[width=0.494\textwidth]{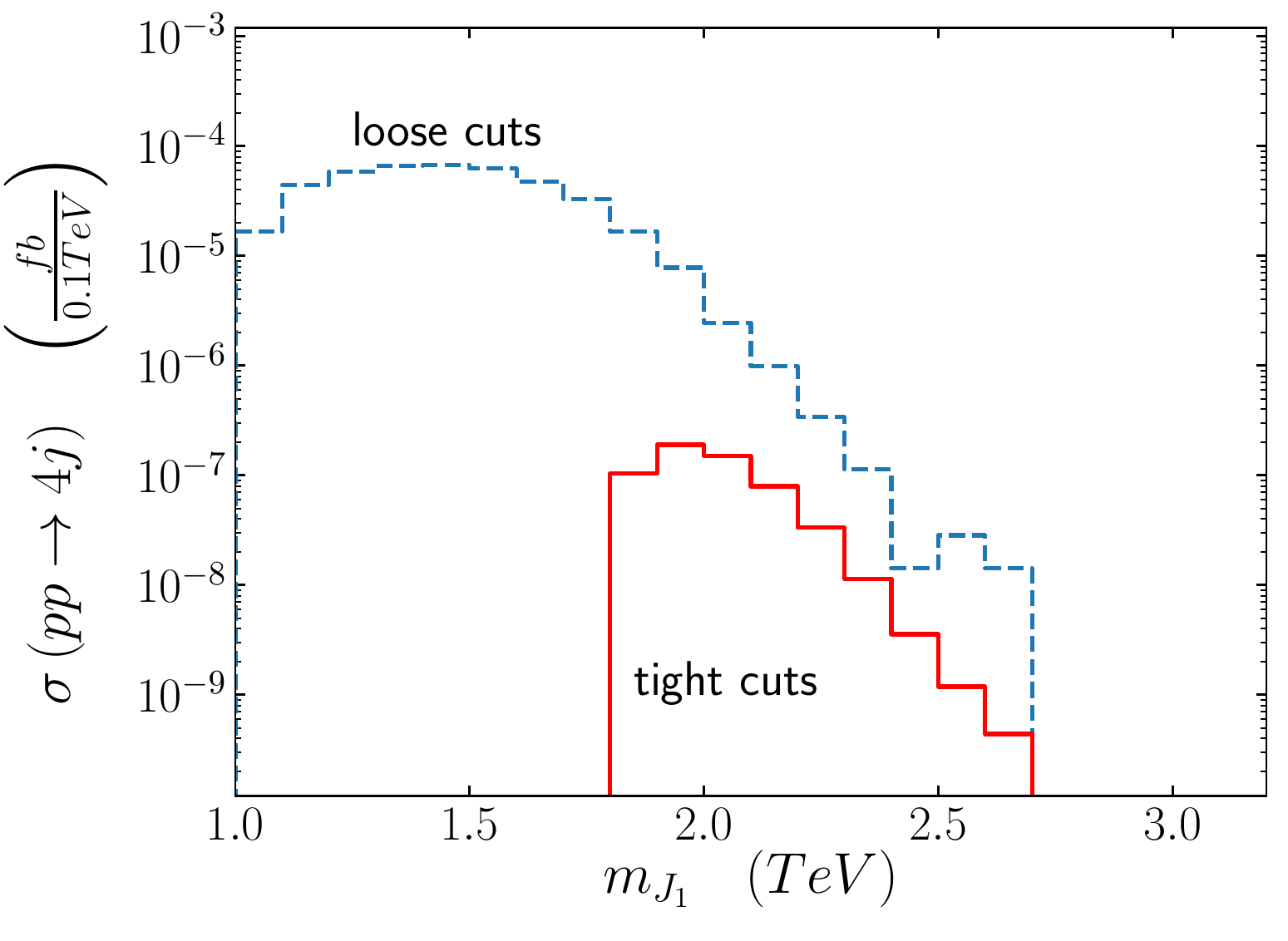}    %eps} 
\hspace*{1mm}
\includegraphics[width=0.487\textwidth]{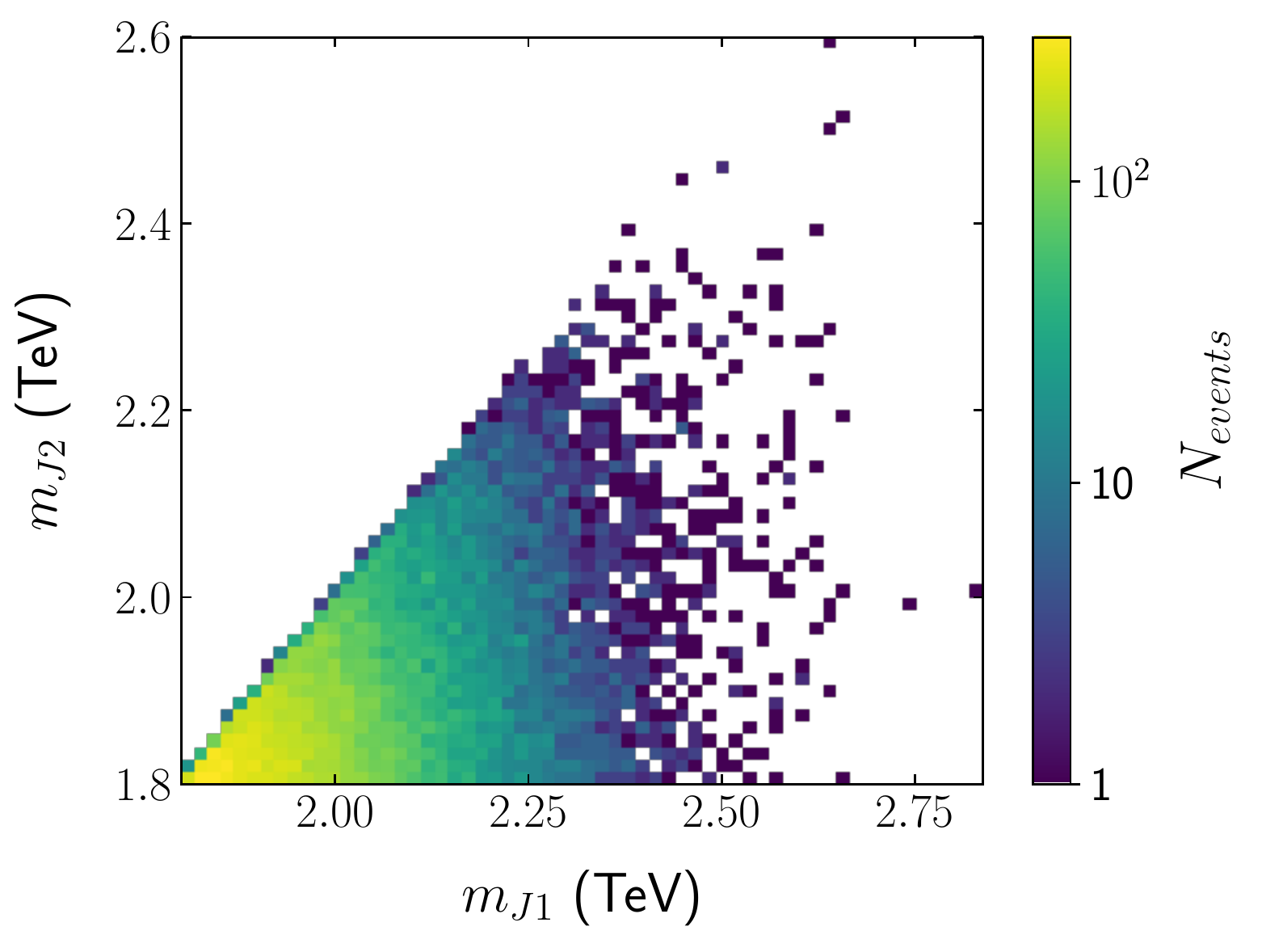}

\vspace*{-4mm}
\caption{{\it Left panel:} same as Figure~\ref{fig:MJ2andM4j} but for the larger wide-jet mass, $m_{J1}$.
{\it Right panel:} the distribution of $m_{J2}$ vs. $m_{J1}$ for the sample with tight cuts (see text). } %\\ [-3mm] }
\label{fig:MJ1andMJ2}
\end{figure}
%%%

%%%
\begin{figure}[t]
\hspace*{-1mm}\includegraphics[width=0.497\textwidth]{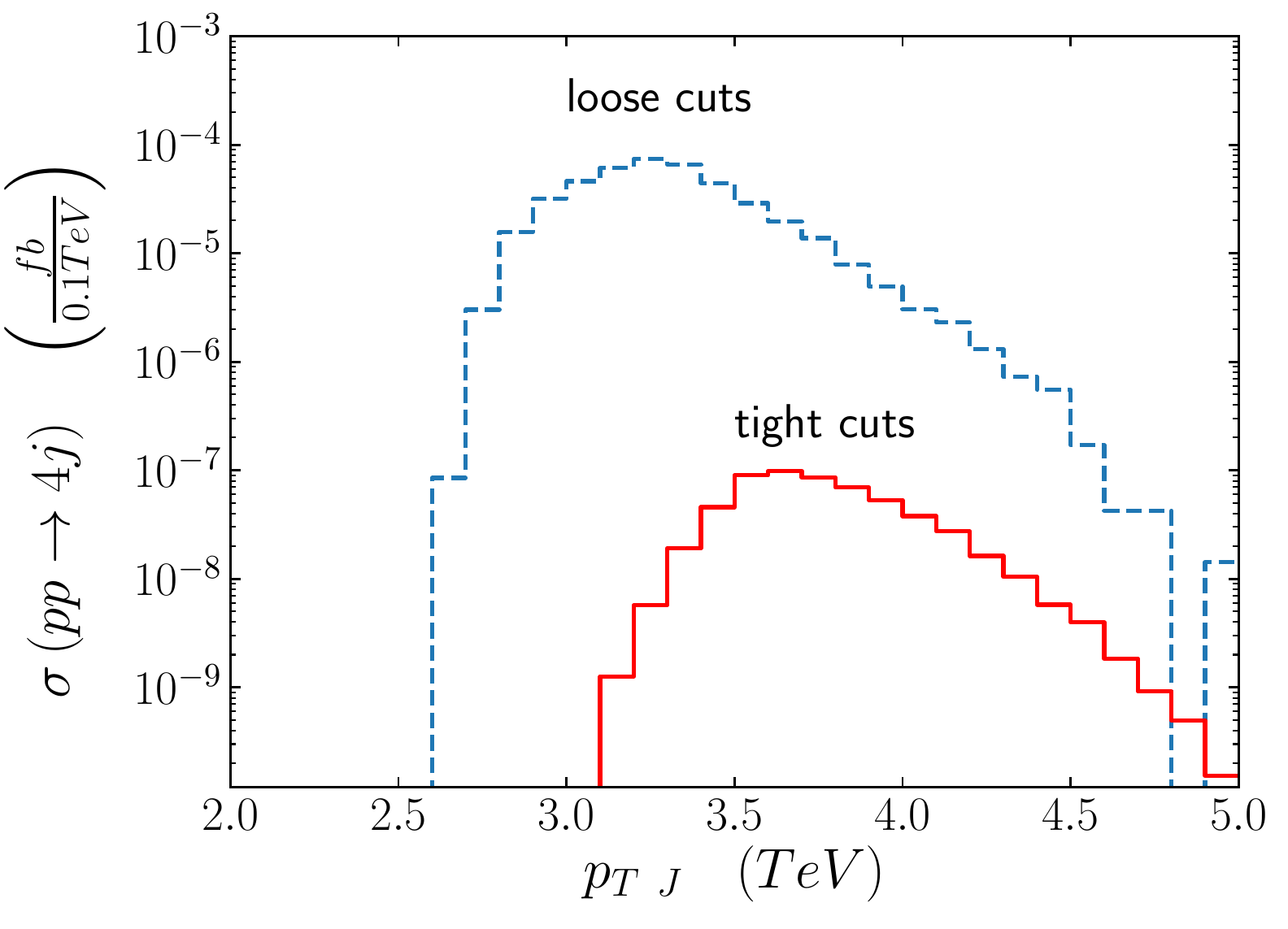}    %eps} 
\hspace*{1mm}
\includegraphics[width=0.489\textwidth]{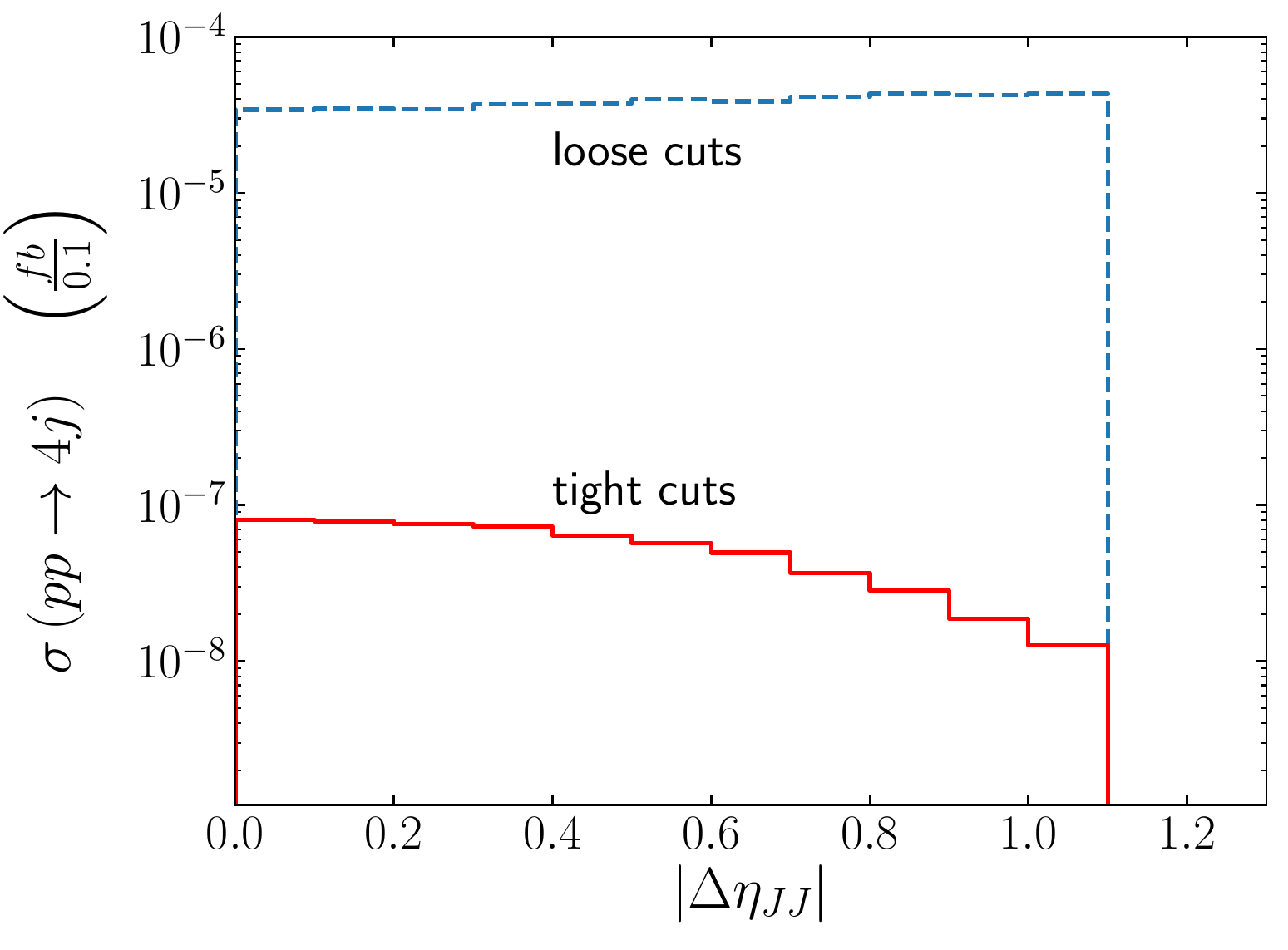}    %eps}

\vspace*{-4mm}
\caption{Same as Figure~\ref{fig:MJ2andM4j} but for the transverse momentum $p_{T_J}$ of the wide jets  (left panel), 
and for the pseudorapidity separation $|\Delta \eta_{JJ}|$ between the wide jets  (right panel).
\\ [-0.2mm]}
\label{fig:PTandEta}
\end{figure}
%%%

In Figure~\ref{fig:PTandEta} we show distributions of the wide jet transverse momentum, $p_{T_J}$, which is the
same for both wide jets, and distributions of the pseudorapidity separation of the two wide jets, $\Delta
\eta_{JJ}$.
Similar to the discussion after Figure~\ref{fig:MJ2andM4j}, the wide jet mass requirements 
$m_{J1}, m_{J2}>1.8$ TeV in the tight sample, combined with the wide jet clustering requirement $\Delta R_J<1.1$, 
constrain the $p_{T_J}$ in the tight sample to be higher than in the loose sample.  

\bigskip
\smallskip

%%%%%%%%%%%%%%%%%%%%%%%%%%%%%%%%%%%%%%%%%%%%%%%%%%%%%%%%%%%%%%%%%%%%%%%%%%%%%%%%%%%%%%%
\subsection{Signals of an 8 TeV particle} 
\label{sec:signals8TeV}

Even though there is a single 4-jet event  reported so far with an invariant mass of 8 TeV, the tiny probability for QCD to 
produce that warrants an investigation of what new particles could produce events of this kind. 
Such an investigation might reveal what additional searches would test the hypothesis that the 8 TeV event is not induced by QCD.

%%%%%%%%%
\subsubsection{$S_{uu}$ diquark with a mass of 8 TeV}

The diquark scalar $S_{uu}$ studied in Section \ref{sec:diquark} is a prime candidate for producing events at very high mass. 
We consider the decay of $S_{uu}$ into two vectorlike quarks, which has a large branching fraction. 
Each vectorlike quark then decays into $ug$, producing two jets, as discussed in section \ref{sec:signalDiquark}.
This leads exactly to the topology of the 8 TeV 4-jet event.
We now study the parameter values that  are consistent with 
this $pp \to S_{uu} \to \chi\chi \to 4j$ process as the origin of the 4-jet event.

The masses of the new particles are given by $M_S = 8$ TeV and $m_\chi = 1.8$ TeV. 
The other important parameters are the couplings $y_{uu}$, which sets the overall production rate, and 
$\sqrt{y_{\chi_R}^2+y_{\chi_L}^2}$, which determines the branching fraction for $S_{uu} \to \chi\chi$.
As explained in Section \ref{sec:signalDiquark}, we take the branching fraction for $\chi \to jj$ to be nearly 100\%.

For simplicity, we set $y_{\chi_L} = 0$ in what follows.
Note that the $U(1)$ flavor symmetry discussed in Section \ref{sec:diquarkInt} can forbid the 
$S_{uu}$ coupling to $\chi_L$ in Eq.~(\ref{eq:diquarkYukChi}),
provided the mass of the vectorlike quark arises from a Yukawa coupling to the $U(1)$ breaking scalar: $\phi_u \bar \chi_L  \chi_R$.
Thus, the only parameters that remain to be fitted to the data are $y_{uu}$ and $y_{\chi_R}$.

For $M_S = 8$ TeV,
the NLO cross section at $\sqrt{s} = 13$ TeV  is 
\be
\sigma_8 (pp \to S_{uu} ) \approx 0.88 \, y_{uu}^2 \; {\rm fb}  ~,
\label{eq:rate8}
\ee
and the $S_{uu} $ branching fraction relevant for the $4j$ final state when $m_\chi = 1.8$ TeV is 
\be
B_8 ( S_{uu} \to \chi\chi) \approx  \left( 1+  1.25\, \frac{ y_{uu}^2}{y_{\chi_R}^2} \right)^{\! \! -1}   ~~,
\label{eq:Br8}
\ee
where the factor of 1.25 is due to the $m_\chi/M_S$ dependent factors in the width for $S_{uu} \to \chi\chi$ given in Eq.~(\ref{eq:SchiWidth}).
Note that the model allows in principle a signal rate much larger than that required to explain the $4j$ event observed in the CMS dijet search. 
For example, $y_{uu} = y_{\chi_R} = 2$ corresponds to $ \Gamma_S/M_S \approx 7 \%$ and
the  $ \sigma_8 (pp \to S_{uu} ) B_8 ( S_{uu} \to \chi\chi) \approx 1.5$ fb, so the expected number of 8 TeV $4j$ events in 77.8 fb$^{-1}$ could have been 
substantially larger than 1.

A  constraint we need to study is the absence of events involving pairs of dijet resonances with a mass of 1.8 TeV in the dedicated searches of that type.
The latest CMS search \cite{Sirunyan:2018rlj} for pairs of dijets included only the 2016 dataset, so it could not observe the 
4-jet event with a pair of dijets at 1.8 TeV recorded in the 2017 dataset  (observed in the dijet resonance search \cite{CMS-PAS-EXO-17-026}). 
Nevertheless, the absence of events at 1.8 TeV in that search sets an upper bound on the cross section for 
$pp \to S_{uu} \to \chi\chi$.
The similar search from ATLAS \cite{Aaboud:2017nmi} is less sensitive to our signal because of a kinematic cut which requires the jets 
within each pair to be highly separated.

Using the MadGraph event generator \cite{Alwall:2014hca}, with model files created by 
FeynRules  \cite{Alloul:2013bka}, we simulated a diquark of mass equal to 8 TeV undergoing 
decays into a pair of vectorlike quarks that produce dijet resonances of 1.8 TeV, and 
we determined the parton-level acceptances  of the CMS search \cite{Sirunyan:2018rlj}  for pairs of dijets. 
The results are in shown in Table~\ref{tab:4JetCutFlow}, where each row includes the combined acceptance for all the requirements above 
and within that row.
We list only those requirements that have less than full acceptance. The requirements 
that each of the four jets has a transverse momentum $p_{Tj}>80$ GeV and a pseudorapidity $|\eta_j|<2.5$ have a  high acceptance.
The requirement that the  mass asymmetry between two pairs of jets is small,
$M_{\rm asym}=(m_{J1}-m_{J2})/(m_{J1}+m_{J2})<0.1$, 
has an acceptance of about 0.9. 
Finally, the requirement on the pseudorapidity separation between the dijet pairs, $|\Delta\eta_{JJ}|<1.0$, 
has a smaller acceptance which depends on the
angular distribution of the model. 

\begin{table}[t]
\begin{center}
\begin{tabular}{|c|c|c|c|}
\hline
\ Pair of dijet    search  \ & \multicolumn{3}{|c|}{Acceptance (\%)}\\  \cline{2-4}
 \hspace*{0.5cm}    requirement     \hspace*{0.5cm}      &  \hspace*{0.5cm} $S_{uu} \to \chi \chi$   \hspace*{0.5cm}  &  \hspace*{0.5cm}   $G' \to \chi \bar \chi$   \hspace*{0.5cm}  &  \hspace*{0.5cm}   $G' \to \Theta \Theta$   \hspace*{0.5cm}  \\ \hline
$p_{Tj}>80$ GeV        &     98.4    &   98.3  &  98.4\\
$|\eta_j|<2.5$        &     94.1    &   93.0  &  96.8\\
$M_{\rm asym}<0.1$          &     80.9    &   83.6  &  89.4\\ 
$|\Delta\eta_{JJ}|<1$ &     40.4    &   36.1  &  59.1\\  \hline
\end{tabular}
\end{center}
\caption{Cut flow for signals of a diquark ($S_{uu}$) or a coloron ($G'$) 
with mass of 8 TeV decaying to pairs of particles with mass of 1.8 TeV, each decaying into two jets.
The percentage of events accepted is shown after 
applying each of the selection requirements from the  CMS search \cite{Sirunyan:2018rlj} for pairs of dijets.  \\ }
\label{tab:4JetCutFlow}
\end{table}

We find that the $S_{uu} $ signal acceptance for the kinematic cuts imposed in the CMS search \cite{Sirunyan:2018rlj} 
is approximately 0.4, as can be seen in Table~\ref{tab:4JetCutFlow}. 
The nonobservation of 4-jet events with two dijet masses near 1.8 TeV implies, based on Poisson statistics,  that the upper limit at the 95\% CL
on the predicted number of signal events is 3. The amount of data analyzed in that CMS search is 35.9 fb$^{-1}$.
Thus, the upper limit on the  cross section  times  branching fraction is 
\be
\sigma_8(pp \to S_{uu} ) \; B_8 ( S_{uu} \to \chi\chi) < 0.21 \; {\rm fb}  ~~.
\label{eq:constraint4j}
\ee
From  Eqs.~(\ref{eq:rate8}) and (\ref{eq:Br8}) it follows that 
the above limit  translates into a constraint on the following combination of diquark couplings:
\be
f(y_{uu},y_{\chi_R}) \equiv y_{uu} \,   \left( 1+  1.25\, \frac{ y_{uu}^2}{y_{\chi_R}^2} \right)^{\! \! -1/2}  < \, 0.49 ~~.
\label{eq:function}
\ee

Let us now turn to the CMS dijet  resonance search  \cite{CMS-PAS-EXO-17-026} that observed the 8 TeV 4-jet event. 
The parton-level signal acceptance of the CMS search for reconstructing a 
resonance of mass equal to 8 TeV decaying to pairs of wide jets is shown in
Table~\ref{tab:DijetCutFlow}.  First, the requirement that each of the four standard jets in the final state satisfies 
$|\eta_j|<2.5$ has
an acceptance above 0.93.  We then apply the CMS wide jet algorithm associating the third and fourth jets to the closest of the two 
leading jets, and require that this association results in two pairs of dijets, rejecting events where there is a cluster of three jets 
within a wide jet. 
The acceptance for this requirement, shown by the row labelled 
``Dijet pairs" in Table~\ref{tab:DijetCutFlow}, is roughly 3/4.  Next we require the two clusters of jets have a pseudorapidity separation 
$|\Delta\eta_{JJ}|<1.1$, which has an acceptance which again depends on the angular distribution. 

\begin{table}[t]
\begin{center}
\begin{tabular}{|c|c|c|c|}
\hline
 \  CMS dijet search \ & \multicolumn{3}{|c|}{Acceptance (\%)}\\ \cline{2-4}
   requirement          &  \hspace*{0.5cm}   $S_{uu} \to \chi \chi$  \hspace*{0.5cm}   &  \hspace*{0.5cm}  $G' \to \chi \bar \chi$   \hspace*{0.5cm}   &  \hspace*{0.5cm}    $G' \to \Theta \Theta$  \hspace*{0.5cm}   \\ \hline
$|\eta_j|<2.5$         &     94.6    &    93.5 & 97.3 \\
Dijet pairs             &     69.2    &    69.0 & 75.3 \\
$|\Delta\eta_{JJ}|<1.1$ &     44.3    &    38.9 & 58.8 \\
$\Delta R_J<1.1$        &      8.3    &     9.4 & 14.0 \\ \hline
\end{tabular}
\end{center}

\vspace*{-2mm}
\caption{Cut flow for signals of ultraheavy resonances decaying to pairs of dijets reconstructed as wide jets. 
The percentage of events accepted after applying each of the selection requirements for an ultraheavy reasonance of 
mass 8 TeV decaying to pairs of particles of mass 1.8 TeV. \\ }
\label{tab:DijetCutFlow}
\end{table}

\begin{figure}[t] \vspace*{-3mm}
\hspace*{-1mm}\includegraphics[width=0.46\textwidth]{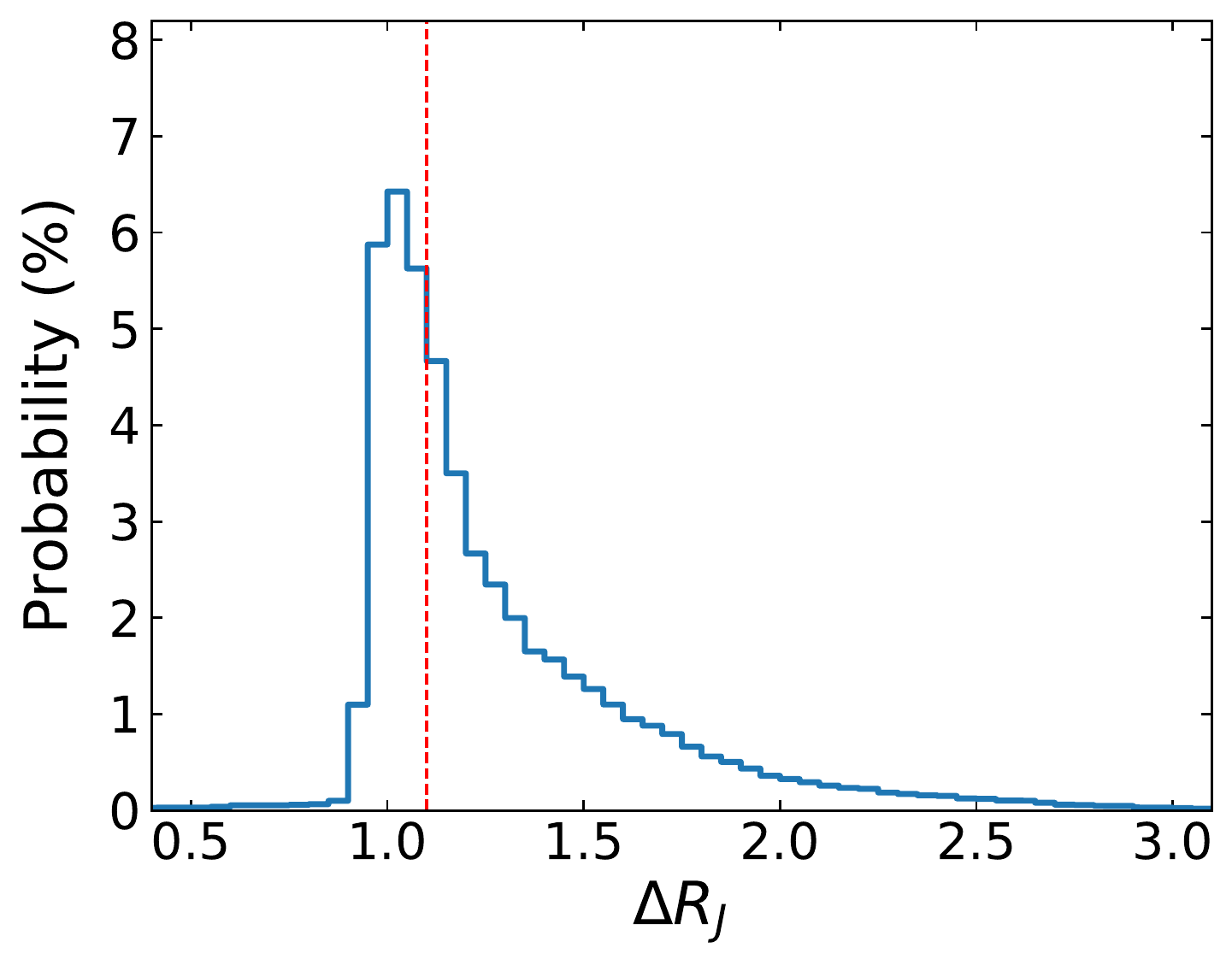}    %eps}   

\vspace*{-5mm}
\caption{The separation $\Delta R_J$ of the two standard jets within a wide jet produced by the $\chi \to jj $ decays
for events originating from  $S_{uu} \to \chi\chi$, 
in the model with a diquark of mass $M_S = 8$ TeV and a vectorlike quark of mass $m_\chi =1.8$ TeV. 
Only the first three selection cuts from Table \ref{tab:DijetCutFlow} have been imposed. The dotted red line denotes the location of the 
upper limit of the  $\Delta R_J$ cut.
\\ [-4mm]}   
\label{fig:deltaR}
\end{figure}

 Finally, we apply the CMS
requirement on the separation between the jets in each wide jet, $\Delta R_J<1.1$, which has an acceptance of about 0.2. 
The reason for this low acceptance can be seen from the $\Delta R_J$ distribution in Figure~\ref{fig:deltaR}. The distribution peaks at 
$\Delta R_J \approx 1.1$ and has a long tail to larger values of $\Delta R_J$, so the wide jet requirement $\Delta R_J<1.1$ is too tight for this 
search. The acceptance can be significantly increased by relaxing this cut or replacing the CMS wide jet algorithm with one optimized for
a 4-jet signal of an ultraheavy resonance.
The final acceptance of the 4-jet analysis (see Table~\ref{tab:4JetCutFlow})  is substantially larger
than the acceptance of the dijet analysis (see Table~\ref{tab:DijetCutFlow}), 
mainly because the $\Delta R_J<1.1$ requirement for wide jets is only applied in the dijet analysis.

The diquark signal can  produce the CMS 4-jet event with a reasonable probability even considering the constraint from the 4-jet search.
Given the acceptance of 8.3\% for the $S_{uu}$ signal and the constraint (\ref{eq:constraint4j}) from the absence of events in the 4-jet search, 
we find that the maximum number of expected 4-jet events in the CMS dijet search with 77.8 fb$^{-1}$ of data 
is $N_{4j}^{\rm exp} \approx 1.3$. In that case the probability for observing at least one event is 73\%.
It is perhaps more illuminating to determine 
what is the maximum probability $P_{\rm max}$ for having at  least one $4j$ event  in the dijet search
and no events in the pair-of-dijets search. We find  
$P_{\rm max} \approx  14\%$, which is  
obtained for $\sigma_8(pp \to S_{uu} ) \, B_8 ( S_{uu} \to \chi\chi) \approx  5.8 \times 10^{-2}$ fb,  
or equivalently for $f(y_{uu},y_{\chi_R}) \approx  0.26$, where $f$ is the function defined in Eq. (\ref{eq:function}).
The above probabilities, 73\% and 14\%, assume no correlation between the results of the 4-jet search and the dijet search.  
If there was complete correlation between the observation of zero events in the 4-jet search and in the 
first 35.9 fb$^{-1}$ of the dijet search, the above probabilities would be reduced to 50\% and 8\%, respectively.

Let us  illustrate the event rates by choosing 
a benchmark in the parameter space: $y_{uu} = 0.3$, $y_{\chi_R} = 0.5$, corresponding to  $f(y_{uu},y_{\chi_R}) = 0.25$.
This implies $B_8 ( S_{uu} \to \chi\chi)  \approx   69\%$ and  $\sigma_8(pp \to S_{uu} ) \approx  8.0  \times 10^{-2}$ fb. 
The expected number of 4-jet signal events is then $N_{4j}^{\rm exp}  \approx  0.34$ in the CMS dijet search,
and $N_{2\times 2j}^{\rm exp}  \approx  0.79$  in the CMS pair of dijets search. 
The expected number of signal $jj$ events in the CMS+ATLAS searches (with a total of 77.8 + 37.0 fb$^{-1}$) is  $N_{jj}^{\rm exp}  \approx  1.4$.
Note that the diquark  is very narrow, with a total width for this bechmark given by $\Gamma_S \approx 23$ GeV, {\it i.e.}, $ \Gamma_S/M_S \approx 0.4 \%$. 

Similar agreement with  the data is obtained for a range of coupling values. For example,  $y_{uu} = 0.35$, $y_{\chi_R} = 0.7$,  corresponding to   $f(y_{uu},y_{\chi_R}) = 0.31$,
gives  $B_8 ( S_{uu} \to \chi\chi)  \approx  76\%$ and  $\sigma_8 (pp \to S_{uu} ) \approx  0.11$ fb, corresponding to 
$N_{4j}^{\rm exp}  \approx  0.52$, $N_{2\times 2j}^{\rm exp}  \approx  1.2$ and $N_{jj}^{\rm exp}  \approx  1.5$.  
We conclude that the model with a scalar diquark and a vectorlike quark can account for all 
observations, and has a probability of fitting  the data which is more than three orders of magnitude higher than in the case of the SM.

To further optimize the search for an 8 TeV resonance decaying to a pair 
of particles
of mass 1.8 TeV at the LHC, we propose a search in the $m_{4j}$ versus
$\overline{m}_{jj}$
plane. The pairs of standard jets should be chosen to minimize the pair 
mass asymmetry, and we find this procedure increases the signal 
acceptance to 100\%
at the parton level for the existing CMS cut $M_{asym}<0.1$. The 
imposition of the $\Delta R_{JJ}$ and $\Delta \eta_{JJ}$ cuts is not 
required to reduce the negligible QCD backgrounds near the interesting 
values of $m_{4j}$ and $\overline{m}_{jj}$.  As both cuts significantly
reduce the acceptance for signal events, these requirements should 
either be dropped or substantially relaxed.

%%%%%%%%%%%%%%%%%%%%%%%%%%%%%%%%%%%%%%%
\subsubsection{Other models}  

The coloron discussed in Section \ref{sec:coloron} may also produce events with the topology of the observed ones at 8 TeV.
The processes $pp \to G' \to \Theta\Theta \to 4j$ (in the coloron+scalar model) or  
$pp \to G' \to \chi \bar \chi \to 4j$  (in the coloron+quark model) may generate the 4-jet event 
for $M_{G'} = 8$ TeV and a color-octet scalar of mass $M_\Theta =1.8$ TeV, or a vectorlike quark of mass $m_\chi =1.8$ TeV.
Dijet events near 8 TeV are also generated by the $pp \to G' \to q\bar q$ process.

To find the rates for these processes, we multiply the coloron production cross section at the 13 TeV LHC  
from Figure \ref{fig:xsec} by the branching fractions from Figure \ref{fig:BR}. The result is shown in Figure \ref{fig:rateGp}, where it is evident that 
the cross section times branching fraction for  the 4-jet final state has a theoretical upper limit: $1.1\times 10^{-3}$ fb for $pp \to G' \to \Theta\Theta$  at $\tan\theta = 0.34$, and 
$3.4 \times 10^{-3}$ fb for $pp \to G' \to \chi \bar \chi$  at $\tan\theta = 0.64$.

\begin{figure}[t]
\vspace*{9mm}
\hspace*{-1mm}\includegraphics[width=0.48\textwidth]{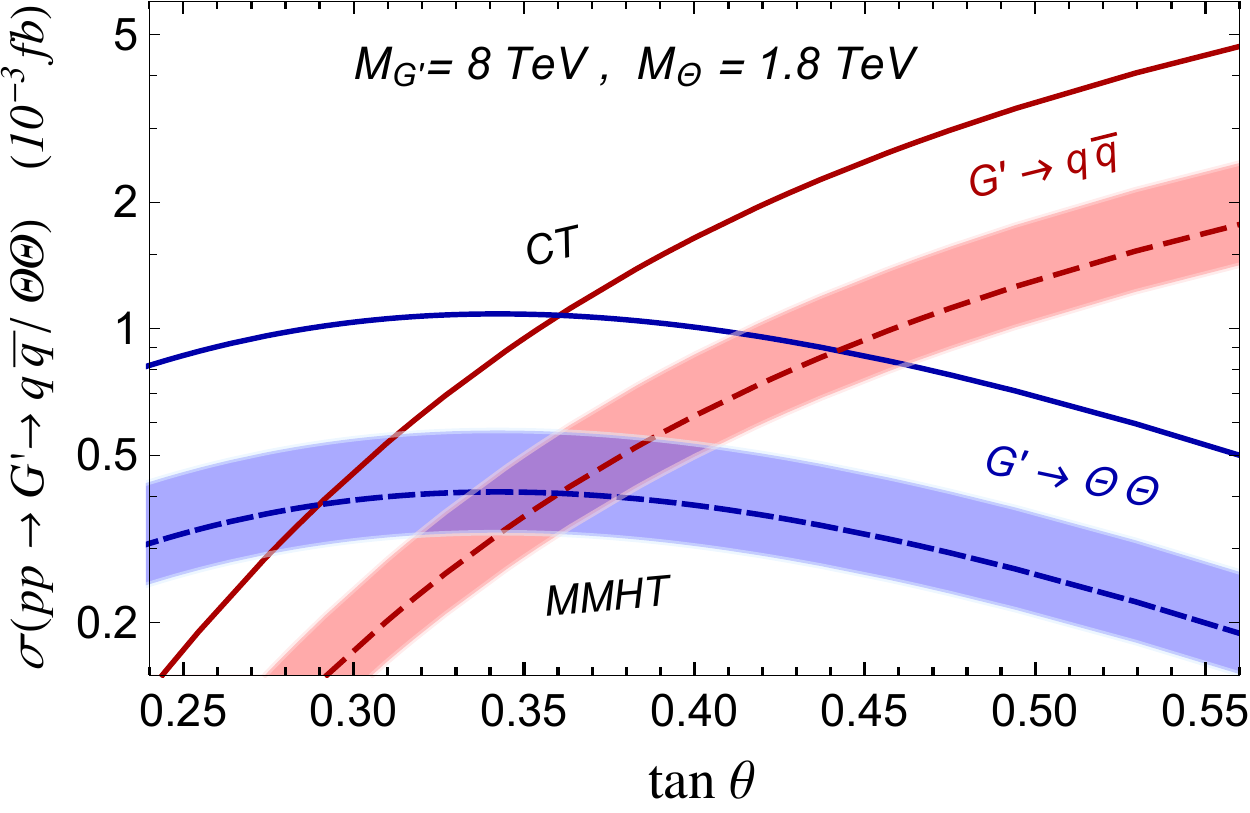}\vspace*{-1mm} \hspace*{3mm}\includegraphics[width=0.48\textwidth]{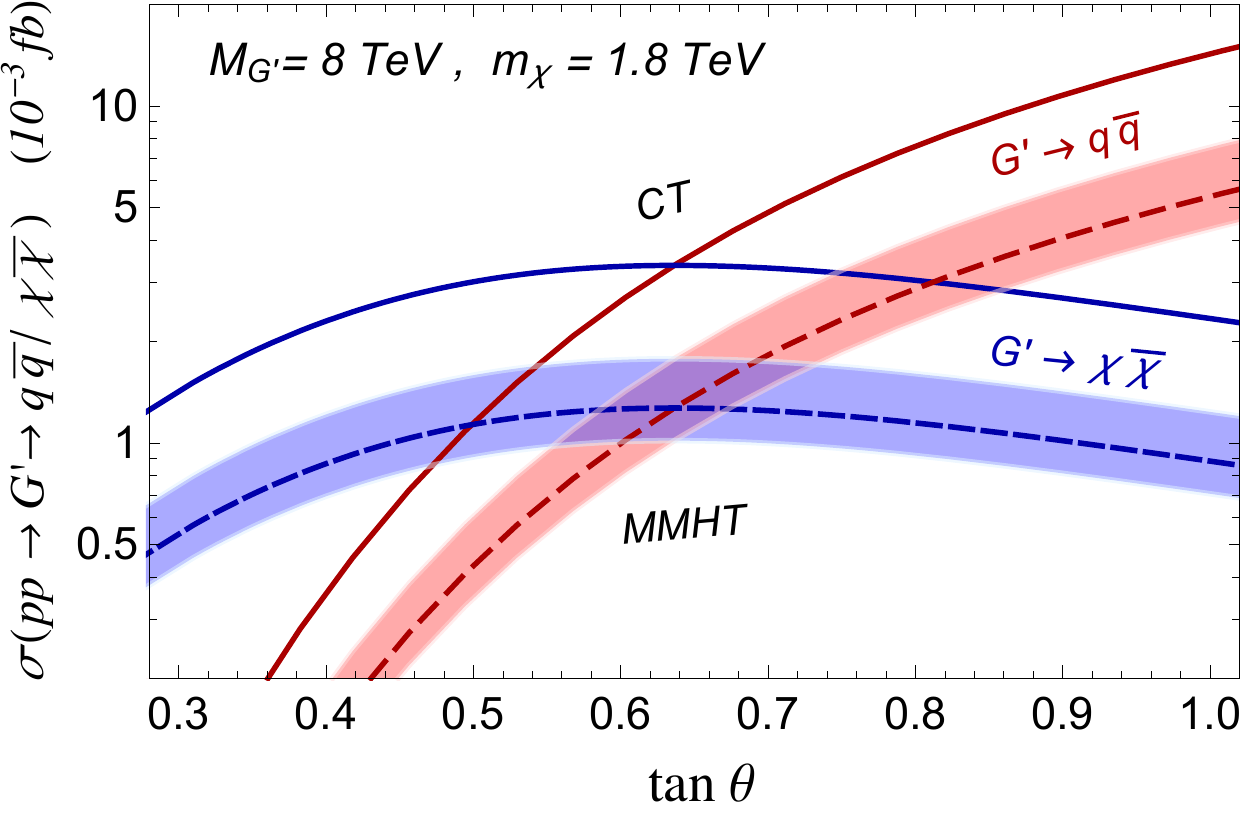}\vspace*{-3mm}
\caption{Cross section times coloron branching fractions at the 13 TeV LHC for $G' \to  \Theta \Theta$  (left panel, blue lines) or $G' \to  \chi \bar \chi $  (right panel, blue lines),
and for $G' \to q\bar q$ summed over six flavors (red lines),    
as a function of  the coupling normalization $\tan\theta$. The computation is at LO 
with $M_{G'} = 8$ TeV and $M_\Theta = 1.8$ TeV (in the coloron+scalar model, left panel) or $m_\chi = 1.8$ TeV
(in the coloron+quark model, right panel). Two PDF sets at LO are used here: 
solid lines correspond to the CT14 set, while the dashed lines and the shaded bands correspond to the central value and 
uncertainty of the MMHT2014 set, respectively. 
 \\ [-3mm] }
\label{fig:rateGp}
\end{figure}

The acceptance for these processes in the CMS dijet analysis is given in Table~\ref{tab:DijetCutFlow}: 14\% in the case of $G' \to \Theta\Theta$,
and  9.4\% in the case of $G' \to  \chi \bar \chi$. 
The maximum cross section times branching fraction, multiplied by the acceptance and the amount of data (77.8 fb$^{-1}$),
then gives the largest expected number of events: $N_{4j}^{\rm exp} \approx 0.012$ in the coloron+scalar model, and $N_{4j}^{\rm exp}  \approx 0.025$ in the coloron+quark model.
Thus, the probability for the 8 TeV 4-jet event to be due to a cascade decay of the coloron is at most 2.5\%.
Despite being a low probability, this is larger by more than two orders of magnitude than the probability predicted in the SM. 
If $pp \to G' \to \chi \bar \chi$   is indeed the origin of the 4-jet event, the probability for observing a second event of this type in 300 fb$^{-1}$ of data at $\sqrt{s} = 13$ TeV 
is at most 9\%. At $\sqrt{s} = 14 $ TeV, the coloron production cross section is larger by a factor of 4.0 when the CT14 PDF set is used, and of 4.5 when the MMHT2014 set is used. 

There is a larger contribution from the $pp \to G' \to q \bar q$  process  to the $jj$ events in the CMS+ATLAS searches, because the 
 acceptance is higher ($\sim 0.4$), and there is more data (77.8 + 37.0 fb$^{-1}$). In practice, though, that contribution is smaller than the  
background.
Even for $\tan\theta = 0.8$, which in  the coloron+quark model saturates the $\Gamma_{G'}/M_{G'} < 7\%$ constraint, 
 the expected number of $jj$ events is  only $N_{jj}^{\rm exp}  = 0.35$.  Note that despite the large acceptances 
 listed in Table \ref{tab:4JetCutFlow},   the limit from the search of a pair of dijets  is rather weak due to the small rates in the
coloron models.

We have focused here only on a couple of renormalizable coloron models. Nevertheless, it appears challenging to construct selfconsistent and viable coloron models with substantial
larger rates while keeping the width-to-mass ratio below $\sim 7\%$. In the end the limited rate is due to the very small PDFs for antiquarks at large $x$.

The diquark model discussed earlier in this Section is not unique. For example, the vectorlike quarks could decay into 
an up quark and a boosted spin-0 particle decaying into $gg$, as mentioned in 
Section \ref{sec:signalColoron}. Effectively,  this would still lead to a 
$pp \to S_{uu} \to \chi\chi \to 4j$ process, which could be the origin of the 8 TeV 4-jet event.

The color-sextet diquark scalar $S_{uu}$ has a particularly large production rate because it couples to two up quarks. Diquarks that couple to an up quark and a down quark
also have large production cross sections. A color-antitriplet scalar coupled to $u_R d_R$ would generically also couple to $Q_L Q_L$, where $Q_L$ is the SM quark doublet of the first generation.  
Limits on this type of diquark are explicitly set in the CMS dijet searches \cite{Sirunyan:2018xlo,CMS-PAS-EXO-17-026}. However, its couplings to quark doublets which are connected by the CKM matrix, make this color-antitriplet scalar more likely to induce large flavor-changing processes.

A color-sextet scalar (labelled by $S_{ud}$) that couples to $u_R d_R$ would not couple to $Q_L Q_L$ due to the antisymmetry of the $SU(2)_W$ indices. 
An $S_{ud}$ with mass of 8 TeV could generate a 4-jet final state if there are two vectorlike quarks; however, these would have to be approximately degenerate in mass, near 1.8 TeV, which would require fine tuning.

\smallskip \smallskip 
%%%%%%%%%%%%%%%%%%%%%%%%%%%%%%%%%%%%%%%%%%%%%%%%%%%%%%%%%%%%%%%%%%%%%%%%%%%%%
%%%%%%%%%%%%%%%%%%%%%%%%%%%%%%%%%%%%%%%%%%%%%%%%%%%%%%%%%%%%%%%%%%%%%%%%%%%%%
\section{Conclusions} 
\label{sec:conc}

How far can the LHC push the energy frontier?  
A more specific version of this question is the following: 
``What are the heaviest particles that could be produced and eventually detected?"
In this article we have shown that an  $S_{uu}$ diquark,
a  color-sextet scalar which couples to two up quarks, can 
be discovered as a narrow dijet resonance, by the end of the high-luminosity run of the LHC, 
if its mass is as large as 11.5 TeV.
For a particle such as a coloron, a color-octet of spin 1 that couples to 
quark-antiquark pairs, the discovery reach is up to 8.5 TeV.

We have shown that both the diquark and the coloron 
may decay into a pair of new particles, each of them then decaying into two jets.
The ensuing 4-jet signal is substantially larger 
than the QCD  background for ultraheavy resonances
(mass above $\sqrt{s}/2$).

We have compared the LHC 4-jet data to models of an ultraheavy resonance.
The CMS 4-jet event is a candidate for a diquark with a mass of 8 TeV decaying to a pair of 
vectorlike quarks with a mass of 1.8 TeV. We have shown that the expected signal 
at those masses could be as large as $1.3$ events in 78 fb$^{-1}$, corresponding to a probability of $73\%$ of observing at least one event. 
There is however some tension with the nonobservation of events in a CMS search for nonresonant pairs of dijets. 
The probability that the diquark produced at least one event in the dijet search and no events in the  pair of dijets search 
can be as large as 14\%. 

The coloron decay into  a pair of vectorlike quarks, or into two color-octet scalars, cannot produce
more than about $2.5 \times 10^{-2}$ events with those masses, making it less likely as 
an explanation of the CMS event. 

The probability for that 4-jet event to be produced by QCD is much smaller, 
approximately $5\times 10^{-5}$.  While the CMS 4-jet event appears unlikely to be due to QCD, aposteriori 
estimates of the probability of observing one event from the tail of a falling background are inherently biased by
the event, so no firm conclusions can be drawn about the significance of a signal unless additional similar events 
are observed.  We have proposed in Section \ref{sec:signals8TeV} a search for 4-jet resonances with the jets grouped in pairs of high-mass dijets, 
which would have a high acceptance for observing the needed additional events to confirm a signal hypothesis.

We have also compared the LHC dijet data to models of an ultraheavy resonance.
In addition to the CMS 4-jet event, ATLAS and CMS have reported three other events with only two high $p_T$ 
jets, which have the reconstructed mass in the range 7.9--8.1 TeV. The QCD background here is more significant,
roughly 1.6 events for the sum of both experiments assuming a mass window of 7.7--8.3 TeV. Nevertheless, this 
leaves room for a signal in the dijet channel, which must be present at some strength if the CMS 4-jet 
event is also a signal. Assuming the three observed events come from a diquark signal of approximately 1.5 events on the
same size QCD background, the branching fractions of the diquark decays would be 70\% to a 
vectorlike quark pair resulting in 4-jet events, and 30\% to two up quarks resulting in dijet
events. For an optimal search for this class of ultraheavy resonance, dijet searches could be 
synchronized with the 4-jet searches, and the combined signal significance could be evaluated.
 
The LHC has the potential to discover an ultraheavy resonance 
with a cross section beyond the QCD background. The forthcoming runs of the LHC should tell us more
about physics at the 10 TeV scale.

\smallskip \smallskip \smallskip\smallskip
\bigskip\bigskip\bigskip\bigskip\bigskip

\newpage
%%%%%%%%%%%%%%%%%%

{\bf Acknowledgments:} \  We thank Niki Saoulidou, Dimitrios Karasavvas, and Melpomeni Diamantopoulou 
for discussions of the CMS dijet  resonance search and the 4-jet event.
We also thank Stefan H\"oche and Stefan Prestel for helpful discussion on the effects of parton showers on the results.
This work was supported by Fermi Research Alliance, LLC, under Contract No. DE-AC02-07CH11359 with the U.S.
Department of Energy, Office of Science, Office of High Energy Physics.

%%%%%%%%%%%%%%%%%%%%%%%%%%%%%%%%%%%%%%%%%%%%%%%%%%%%%%%%%%%%%%%%%%%%%%%%%%%%%%%%%%%%%%% 

\end{document}